\documentclass[mlhc,table,nosubfloats]{jmlr_mlhc}

\jmlryear{2018}
\jmlrworkshop{Machine Learning for Health Care 2018}

\usepackage{amsfonts}       
\usepackage{bbm}
\usepackage{caption}
\usepackage{subcaption}
\usepackage{multirow}
\usepackage{makecell} 

\newcommand{\ie}{i.e., }

\newcommand{\eg}{e.g., }

\title[Phenotyping Endometriosis]{Phenotyping Endometriosis through Mixed Membership Models of Self-Tracking Data}

\author{\Name{I\~{n}igo Urteaga} \Email{inigo.urteaga@columbia.edu} \\
	\addr Department of Applied Mathemathics\\
	Columbia University, New York, NY, USA
	\AND
	\Name{Mollie McKillop} \Email{mm4234@cumc.columbia.edu}\\
	\addr Department of Biomedical Informatics \\
	Columbia University, New York, NY, USA
	\AND
	\Name{Sharon Lipsky-Gorman} \Email{srg2128@cumc.columbia.edu} \\
	\addr Department of Biomedical Informatics\\
	Columbia University, New York, NY, USA
	\AND
	\Name{No{\'e}mie Elhadad} \Email{noemie.elhadad@columbia.edu}\\
	\addr Department of Biomedical Informatics\\
	Columbia University, New York, NY, USA}

\begin{document}

\maketitle

\begin{abstract}
We investigate the use of self-tracking data and unsupervised mixed-membership
models to phenotype endometriosis. Endometriosis is a systemic, chronic
condition of women in reproductive age and, at the same time, a highly
enigmatic condition with no known biomarkers to monitor its progression and no
established staging. We leverage data collected through a self-tracking app in
an observational research study of over 2,800 women with endometriosis
tracking their condition over a year and a half (456,900 observations overall).
We extend a classical mixed-membership model to accommodate the idiosyncrasies
of the data at hand (\ie the multimodality of the tracked variables). Our
experiments show that our approach identifies potential subtypes that 
are robust in terms of biases of self-tracked data (e.g., wide
variations in tracking frequency amongst participants), as well as to variations
in hyperparameters of the model. Jointly modeling a wide range of observations
about participants (symptoms, quality of life, treatments) yields
clinically meaningful subtypes that both validate what is already known about
endometriosis and suggest new findings.

\end{abstract}

\section{Introduction}
\label{sec:intro}

Smartphones and mobile applications are a powerful way to connect medical
researchers to individuals. Recent software platforms like Researchkit and
ResearchStack facilitate the use of mobile technology to recruit and consent
patients into studies.  The first wave of app-based studies shows that with
the right engagement techniques, patients provide valuable
data through their phone that can, shedding new insight into diseases~\cite{bot2016mpower,chan2017asthma}. This work contributes to the emerging area of research on digital
phenotyping from patient-generated data, specifically from data collected
through smartphone applications~\citep{hartsell2017preliminary,torous2018characterizing,zhan2018using}.

We focus on endometriosis, a condition for which there are no known biomarkers
to help its diagnosis or monitor its progression, and for which subtypes
proposed in the literature are not well established. Using self-tracking data
collected through an app designed specifically for the sake of characterizing
endometriosis at scale, we explore the use of unsupervised methods to identify
subtypes of the disease that cluster patients based on their signs and
symptoms, quality of life, and treatments.

Phenotypes are important as a first step towards a better understanding of endometriosis, so that better treatment and management of patients can be achieved via phenotype-based characteristics. We validate our approach through likelihood evaluations of the model in unseen data, clinical interpretability of identified subtypes by endometriosis experts, purity assessment on a subset
of patient phenotype assignments against clinical experts clustering, and
hypothesis testing against a clinically validated standard questionnaire for
patients with endometriois.

Our experiments show that \textbf{(1)} our approach identifies potential subtypes that
are robust in terms of biases of self-tracked data (e.g., wide
variations in tracking frequency amongst participants), as well as to variations
in hyperparameters of the model; and \textbf{(2)} modeling a wide range of observations
about participants (symptoms, quality of life, treatments) jointly yields
clinically meaningful subtypes that both validate what is already known about
endometriosis and suggest new findings. 

Endometriosis is a chronic condition estimated to affect 10\% of women in
reproductive age~\citep{wheeler1989epidemiology}. It is traditionally
described as when tissue similar to the endometrium (lining of the uterus)
grows outside the uterine cavity, and forms lesions in the pelvic and
gastro-intestinal areas primarily. It impacts women at a systemic
level~\citep{kvaskoff2015endometriosis} and presents with a heavy burden of 
disease~\citep{simoens2012burden}. 

Despite its high prevalence, endometriosis continues to be a highly enigmatic condition: there is lack of specificity in the range of signs and symptoms of the disease, and its clinical characterization is poor~\citep{j-Vercellini2007}. It is understood to be
heterogeneous in nature, and several stages of the disease have been
proposed in the literature. Nevertheless, none correlate with severity of symptoms
experienced by patients, nor do they explain the diversity of symptoms
experienced. In fact, subtyping  of endometriosis is currently an open and important 
research question for the endometriosis community~\citep{johnson2017world}.

There is a need for more accurate phenotyping of endometriosis so that better, more targeted treatments
and management strategies can be developed. However, because endometriosis is not well understood from the cinical point of view, traditional phenotyping approaches
that leverage electronic health record data are not appropriate. We instead
turn to patient-generated data towards that goal.
\paragraph{Technical Significance:}
We contribute to the emerging area of research on unsupervised digital phenotyping from patient-generated data, specifically, by extending mixed-membership models for multi-modal data collected through smartphone applications. 
We show that by jointly modeling multiple types of self-tracking variables, we identify potential disease subtypes
that are robust in terms of biases of self-tracked data, and to variations in model hyperparameters.
\paragraph{Clinical Relevance:}
Endometriosis phenotypes are necessary for a better understanding of the disease, which will lead towards better treatment and management of patients. The proposed unsupervised approach produces clinically relevant groupings of endometriosis signs and symptoms. These phenotypes, grouped by the severity of the condition, suggest novel findings about the disease,
as clinically meaningful associations were identified.

\section{Data and Materials}
\label{sec:material}

The data and materials used in this work are described next. All procedures
were reviewed and approved by our institutional review board under protocol number AAAQ9812.

\subsection{Phendo: A Smartphone App to Self-Track Endometriosis}
\label{ssec:app}

Using participatory design, we designed and developed Phendo, a smartphone app for
women with endometriosis to self-track their condition \citep{mckillop2016exploring,mckillop2018designing}. The research app is available for iOS~\footnote{Available at \href{https://itunes.apple.com/us/app/phendo/id1145512423?mt=8}{https://itunes.apple.com/us/app/phendo/id1145512423}} and
Android~\footnote{Available at \href{https://play.google.com/store/apps/details?id=com.appliedinformaticsinc.phendo}{https://play.google.com/store/apps/details?id=com.appliedinformaticsinc.phendo}} phones. Participants were recruited through patient advocacy groups and, once enrolled, can self-track a variety of variables.

\begin{figure}[h]
        \centering
        \includegraphics[width=0.85\textwidth]{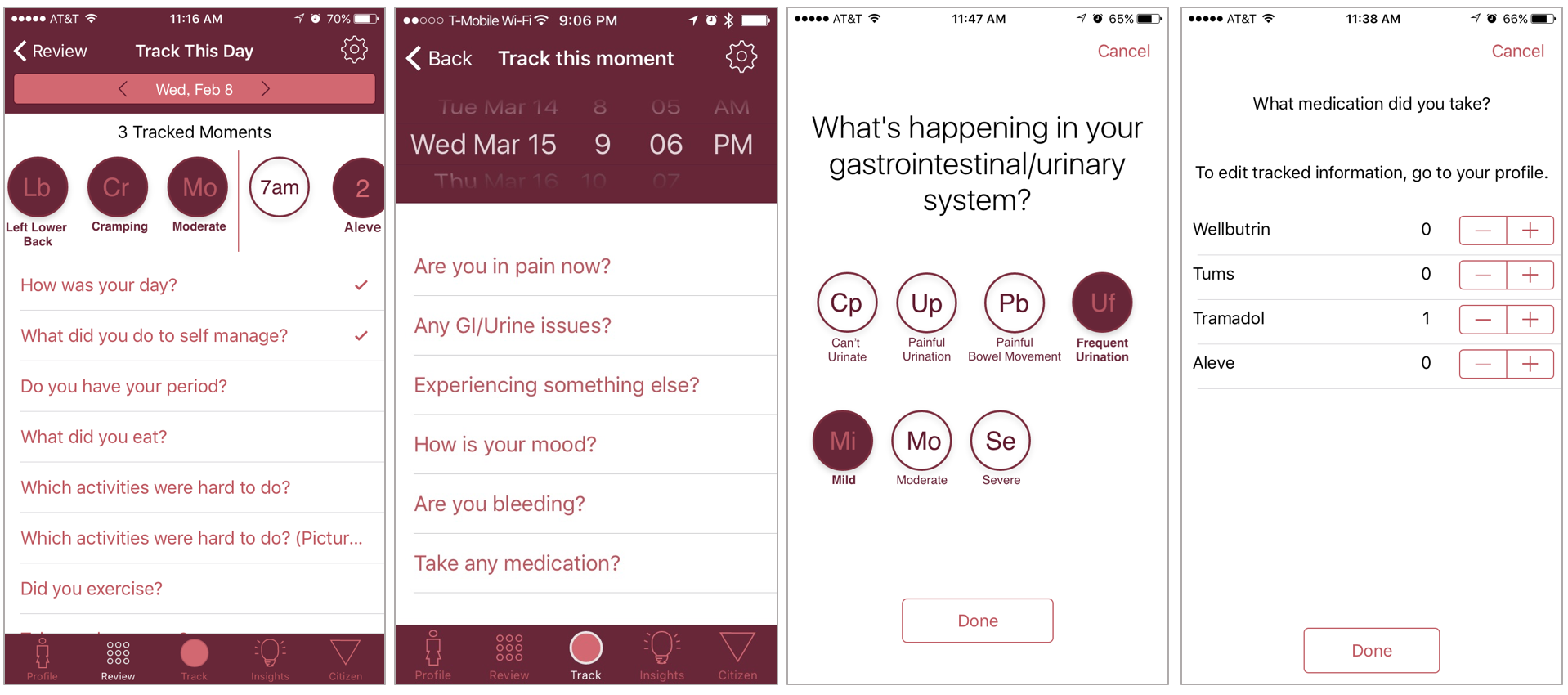}
    \caption{Example screenshots of Phendo, the self-tracking endometriosis app.}    \label{fig:app}
\vspace*{-0.5cm}
\end{figure}

At the \textit{moment} level (i.e., as many times in the day as participants desire), they
can track their pain across 39 specific body locations with 15 modifiers (e.g.,
``cramping" or ``twisting") and 3 severity levels; any of the 15
gastro-intestinal and genito-urinary issues identified during design work with 3
severity levels;  21 signs and symptoms commonly identified by our participants
(e.g., ``blurry vision", ``hot flashes", ``fatigue") and their severity; 10
positive mood and affects and 14 negative ones, 3 bleeding patterns (``clots",
``breakthrough bleeding", ``spotting"), and customized medication intake (see for instance Figure~\ref{fig:app}, where 2 Aleves were tracked at 7am, and other medications appear in the customized medication tracking screen).

At the \textit{day} level, users can track a functional assessment of their day (see question ``How was your day?" in Figure~\ref{fig:app}) from ``great"
to ``unbearable", which activities of daily living were hard to do (customized
for endometriosis needs), menstruation patterns, customized answers for diet
items they want to keep track of, customized supplements, customized exercises,
customized hormonal treatments, sexual activity and potential dyspareunia, and
a daily journal. 

\subsection{Gold-standard data} 
\label{ssec:gold_standard_data}
As part of the profile tab in the Phendo research app, participants
can take a standardized questionnaire designed by the endometriosis research
community, called WERF EPHect~\citep{j-Vitonis2014}.
The questionnaire represents the gold-standard for clinical characterization of endometriosis. It contains information about medical and surgical history, as well as quality of life related questions. 

\subsection{Data preprocessing and cohort selection}
\label{ssec:data_preprocessing_and_cohort}
The data collected for this study contains time-stamped responses from participants about all the variables tracked in the app. Because the data was collected for a wide range of variables at the participants' discretion, it is heterogeneous both in type and quantity across participants, and possibly within a given participant's timeline as well. As a first step towards investigating phenotyping of endometriosis, we ignore the temporal aspect of the condition, and rather aggregate all observations per variables tracked in the app for each participant.

The following variables from Phendo were included in this study (descriptive self-tracking statistics are provided in \autoref{tab:questions_info}): (1) pain location, (2) pain description, (3) pain severity, (4) gastrointestinal and genitourinary (GI/GU) symptoms, (5) their severity, (6) other symptoms, (7) their severity, (8) period flow, (9) bleeding patterns, (10) sexual activity, (11) difficult daily living activities, (12) medications including hormonal treatments, and (13) quality of life (``How was your day?"). 

\begin{table}[!h]
	\begin{center}
		\resizebox{0.99\textwidth}{!}{
			\begin{tabular}{*{3}{|c}|}
				\hline
				Question \cellcolor[gray]{0.6} & \makecell{Number of observations \\ (median/mean/95\%percentile/max)} \cellcolor[gray]{0.6} & \makecell{Number of tracked days \\ (median/mean/95\%percentile/max)} \cellcolor[gray]{0.6}\\ \hline 
				Where is the pain & 4/29/120/2196 & 1/6/31/204 \\ \hline 
				Describe the pain & 4/27/115/1657 & 1/6/31/204 \\ \hline 
				How severe is the pain? & 2/9/43/527 & 1/6/31/204 \\ \hline 
				What are you experiencing & 1/7/33/544 & 1/4/17/175 \\ \hline 
				How severe is the symptom & 1/5/22/544 & 1/4/17/175 \\ \hline 
				Describe the flow & 0/4/16/176 & 0/4/16/176 \\ \hline 
				What kind of bleeding & 0/3/15/173 & 0/2/12/77 \\ \hline 
				Describe GI/GU system & 1/6/28/317 & 1/5/20/205 \\ \hline 
				How severe is it & 1/6/27/317 & 1/5/20/205 \\ \hline 
				Describe sex & 0/1/3/159 & 0/1/3/158 \\ \hline 
				Activities & 7/36/151/2148 & 2/7/29/266 \\ \hline 
				How was your day? & 3/12/57/458 & 3/12/57/458 \\ \hline 
				Medications/hormones taken & 2/14/60/800 & 2/9/37/356 \\ \hline 
				Total & 35/159/718/6065 & 19/71/327/458 \\ \hline 
			\end{tabular}
		}
		\caption{Summary statistics per-tracked question.}
		\label{tab:questions_info}
	\end{center}
\end{table}

\vspace*{-2ex}
We selected a cohort of participants who had self-reported diagnosis of endometriosis, and had at least one entry in one of the above questions between December 2016
(launch of the app) and March 2018, resulting in 2,872 participants (corresponding to 456,900 observations total). Among them, 648 had responded to the WERF EPHect questionnaire.

The app provides a fixed set of possible responses to most of the questions (details about all per-question vocabularies are provided in the appendix). Medications and hormones, which are inputted as free text in the app, were mapped to a fixed size vocabulary by identifying their medication classes.
The WERF EPHect questionnaire responses contained binary (yes/no), categorical (\eg number of laparoscopies) and real numbers (\eg weight) answer types. Not all questions were answered by all selected participants. We only use questions with sufficient number of responses when correlating participants in obtained phenotypes with their responses.

\section{Methods}
\label{sec:method}

There are several challenges to address when using self-tracking data
for phenotyping an enigmatic condition. First, because endometriosis
is a heterogeneous condition and its clinical characterization is poor, no gold-standard phenotypes exist. As such, unsupervised methods are
well-suited to the task of learning phenotypes. Second, self-tracking data is heterogeneous (because of the many variables available for
self-tracking), inherently noisy, irregularly spaced due to
lack of participant engagement, and may reflect biases of self-tracking (e.g., participants may
track more often when they experience symptoms of disease or participants may tend not to self-track when they are sick). To account for these issues, we leverage unsupervised probabilistic methods, and specifically mixed-membership models.

Mixed-membership models are Bayesian generative models used to capture the
latent structure of collections of groups of data. Topic models are their primary example \citep{j-Blei2012}, where one is interested in inferring the latent topics of a corpora of documents. Topic models analyze the statistics of observed words in each document, to capture what the topics are, and what is each document's proportion of topics \citep{j-Blei2003}. 

In this work, we cast phenotyping endometriosis with self-tracking data as a probabilistic topic modeling problem, by considering the set of responses per participant as ``documents", all coming from the ``corpus" of endometriosis patients. As such, each set of observations
is modeled as a mixture model, where the mixture components (the phenotypes)
are shared across the population, but the mixture proportions (the phenotypic profile) vary per participant.
The mixed-membership model infers phenotypes based on the
co-occurrence of observations across the studied set of participants. That is,
``topics" produced by this technique are groupings of responses to self-tracked
variables that describe endometriosis phenotypes. These models are flexible enough to describe participants with more than one of these topics (\ie mixture of phenotypes).

The available self-tracked data however is not a standard document, but a collection of responses to different questions. Therefore, we extend the mixed-membership model to accommodate for multi-modal data, in a similar way done by \cite{j-Pivovarov2015} for EHR phenotyping. In our case, each modality
is an specific question $q=1,\cdots, Q$, with its vocabulary size $V_q$ (see
section \ref{ssec:data_preprocessing_and_cohort} for details). We note that subjects are free to
track whatever questions they want over time and thus, data is highly unbalanced across participants and tracked variables.

The per-question mixed-membership generative process for each subject $s=1,\cdots, S$, follows:
\begin{enumerate}
        \item Draw per-subject phenotypic proportions $\phi_s \sim \operatorname{Dirichlet}_K(\phi|\alpha)$ of dimension $K$ with hyperparameter $\alpha$.
        \item Draw per-phenotype and per-question response proportions $\theta_{k,q} \sim \operatorname{Dirichlet}_{V_q}(\theta|\beta_{k,q})$, for all phenotypes $k=1, \cdots, K$, and questions $q=1,\cdots, Q$, with vocabulary size $V_q$ and hyperparameters $\beta_{k,q}$.
        \item Draw per-subject observation phenotype assignments $z_{s,n} \sim \operatorname{Categorical}_K(z|\phi_s)$, for $n=1, \cdots, N_s$.
        \item Draw per-subject questions responses $x_{s,n}|z_{s,n},q_{s,n} \sim \operatorname{Categorical}_{V_{q_{s,n}}}(x|\theta_{z_{s,n},q_{s,n}})$, for $n=1, \cdots, N_s$, where $q_{s,n}$ indicates the response $n$ to question $q$ by subject $s$.
\end{enumerate}

After observing a dataset with $N_{sq}$ responses per-subject and question, the goal is to infer the phenotypic proportions $\phi_s$ for each participant, and the set of $K$ phenotypes of the disease, parameterized per-question by $\theta_{k,q}$.

To that end, and due to the conjugacy assumptions in the generative process, we resort to a collapsed Gibbs sampler that utilizes the following distribution for observation $n^*$, given $N$ previously ``seen" data points

\begin{equation}
\begin{split}
p(z_{s,n^*}=k|x_{s,n*},q_{s,n*},\alpha_N,\beta_N) &\propto p(z_{s,n^*}=k|\alpha_{s,N}) p(x_{s,n^*}|q_{s,n^*},z_{s,n^*},\beta_{k,q,N}) \; ,\\
\text{with} &\begin{cases} p(z_{s,n^*}=k|\alpha_{s,N}) = \frac{\alpha_{k,s,N}}{\sum_{k=1}^K \alpha_{k,s,N}} \; ,\\
p(x_{s,n^*}|q_{s,n^*},z_{s,n^*},\beta_{k,q,N}) = \frac{\beta_{k,q,v_q,N}}{\sum_{v_q=1}^{V_q}\beta_{k,q,v_q,N}} \; ,
\end{cases} 
\end{split}
\end{equation}

where the parameters for the updated posteriors are of the form
\begin{equation}
\begin{split}
p(\phi_s|Z_N,\alpha_0)&=\operatorname{Dirichlet}_K(\phi_s|\alpha_{s,N}) \; , \qquad \text{with} \; \alpha_{k,s,N}=\alpha_{k,0}+N_{s,k} \;, \\
p(\theta_{k,q}|X_N, Q_N,Z_N, \beta_{k,q,0}) &= \operatorname{Dirichlet}(\theta_{k,q}|\beta_{k,q,N}) \;, \qquad \text{with} \; \beta_{k,q,v_q,N}=\beta_{k,q,v_q,0}+N_{k,q,v_q} \; .
\end{split}
\end{equation}

Note that for the unsupervised learning of phenotypes, we only consider the
self-tracked data, and leave the WERF EPHect questionnaire data for evaluation purposes.

\section{Evaluation} 
\label{sec:evaluation}

\subsection{Experimental Setup}
\label{ssec:approach}
We evaluate and validate the proposed method from different perspectives.

\paragraph*{Likelihood of the learned model on unseen data.} The quality of the
proposed model (and chosen hyperparameters) is evaluated by held-out data
log-likelihood comparisons. We split our dataset into 80/20 train/test splits,
and evaluate the model learned in the training set with its log-likelihood in
the test dataset. We note that computing the log-likelihood of mixed-membership
models is nontrivial, as discussed in \cite{ip-Wallach2009}. The results
presented here are based on extending the ``left-to-right" method proposed by
\citep{ip-Wallach2009} to the per-question mixed membership model of
\autoref{sec:method}.

\paragraph*{Clinical interpretability.} The enigmatic nature of
endometriosis and its poor clinical characterization makes indispensable the
interpretability of the model. We want to understand the specificity of signs
and symptoms of the disease for each learned phenotype, as they are likely to
be heterogeneous. For interpretability, we focus our analysis on per-phenotype
variable posteriors. These per-question posteriors reflect not only which
responses are more commonly tracked per phenotype, but also how they correlate
with each other.

To allow for easy and visually appealing clinical evaluation,
we provide both raw posterior heatmaps and most salient response wordclouds.
Due to the different support size for each considered question, we plot
wordclouds conditioned on the vocabulary items that cover 80\% of the posterior
mass. The maximum font size is fixed equal for all phenotypes within a
question, and the relative size of responses follows the proportions of the
conditional probability ratios. This allows for a more clear identification of
the most salient responses and a principled way of comparing different
sized-vocabularies. Both the heatmaps and word clouds of the posterior variables per phenotypes
were provided to endometriosis experts for review. 

\paragraph*{Agreement between expert clustering and phenotyping.} We randomly selected 30 participants from the cohort, who had at least 30 days of tracked data and had high posterior
probability (above 95\% percent) of being assigned to a unique phenotype (10
participants per phenotype) in the learned model. Note that participants to be evaluated by clinical experts were selected \textit{randomly} within each subtype, based on the model's output only, and that experts were not involved in this selection process at all.
The responses collected from these patients were reviewed by two clinical experts, who were asked to group them into three clusters based on their clinical understanding of patient signs and symptoms. The assignments by the experts and the model are compared, via confusion matrices
and cluster purity metrics. 

\paragraph*{Associations with standardized questionnaire.} We selected a subset of questions from the WERF EPHect questionnaire for validation purposes, which included responses about family history of endometriosis, pelvic pain, menstrual characteristics, surgical procedures, comorbidities, and activities of daily living, as well as general health indicators. We perform association tests between participants assigned to each phenotype (hard clustering based on learned per-participant phenotypic posteriors) and their responses to the questions of interest, and report correlations that are significant. Specifically, categorical question answers were collapsed to `yes' or `no' outcomes for each cluster, producing $2\times2$ contingency tables.  Fisher's exact test \citep{j-Fisher1922} was used for each contingency table to produce an odds ratio, which represents the likelihood of saying `yes' to a particular question and being in a particular cluster, versus the odds of saying `no' to that question and not being in a particular cluster. The average for answers with continuous outcome questions was computed for each cluster, and compared to the mean of those not in that particular cluster, using Welch’s t-test for unequal variances \citep{j-Ruxton2006}. Significance for Fisher's exact tests and Welch’s t-tests was determined at the $0.05$ level.   

\subsection{Results}

\begin{figure}[!h]
	\centering
	\includegraphics[width=0.7\textwidth]{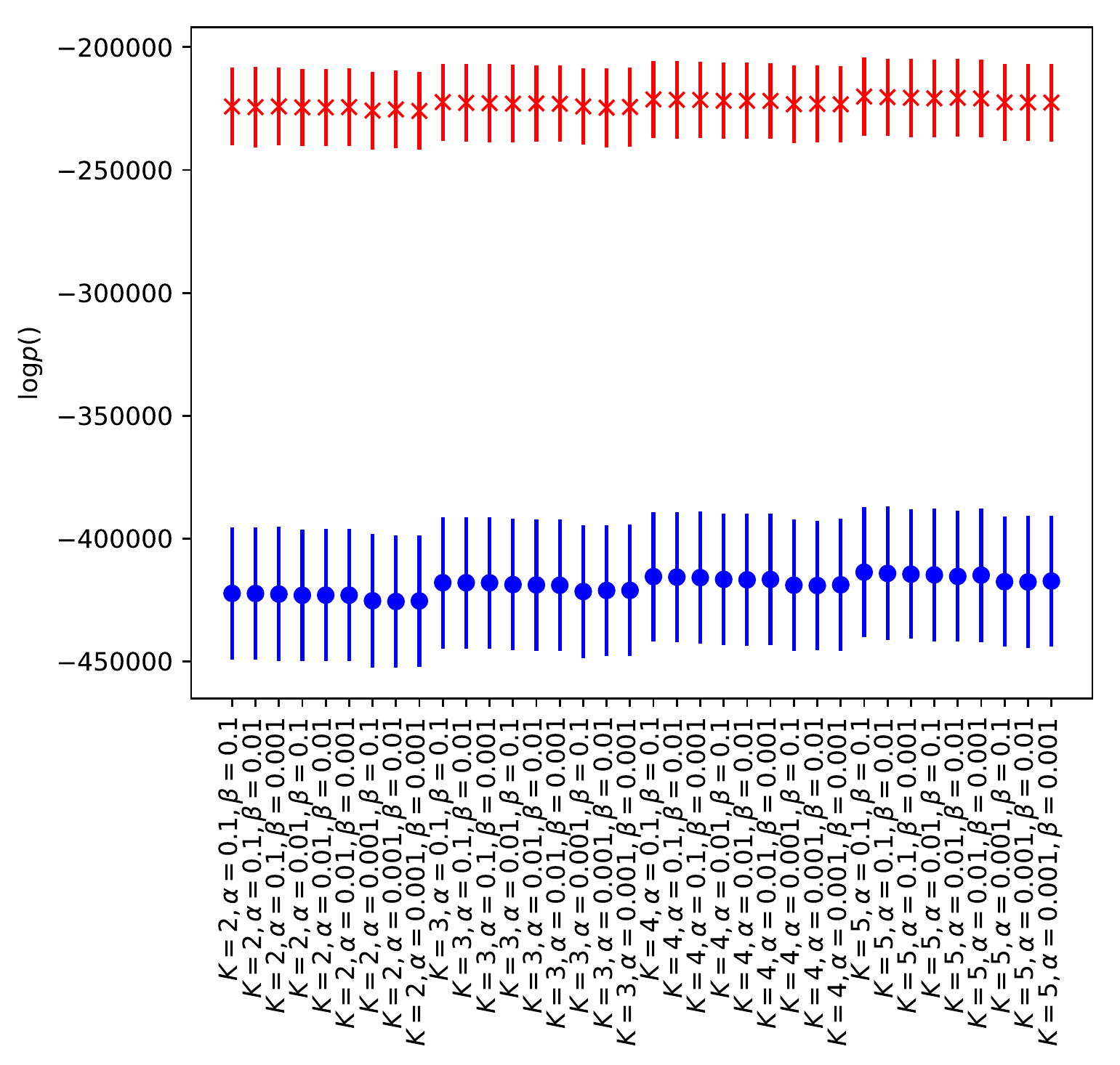}
	\caption{5-fold test data log-likelihood of the proposed method (in red) Vs LDA (in blue).}    
	\label{fig:test_logliks}
\end{figure}

\begin{figure*}[!h]
	\centering
	\begin{subfigure}[h]{0.34\textwidth}
		\includegraphics[width=\textwidth]{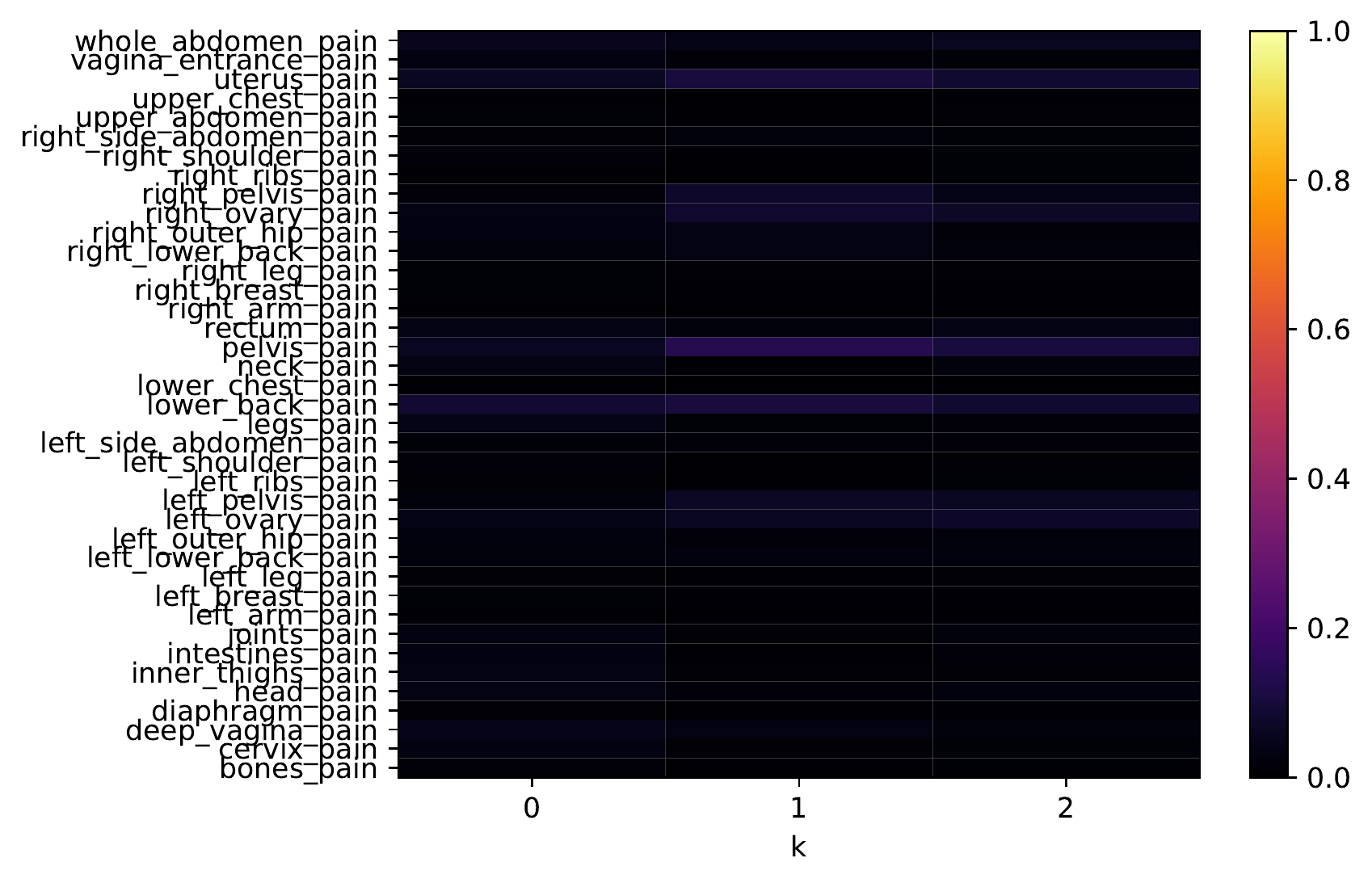}
		\caption{Where is the pain.}
	\end{subfigure}
	\begin{subfigure}[h]{0.30\textwidth}
		\includegraphics[width=\textwidth]{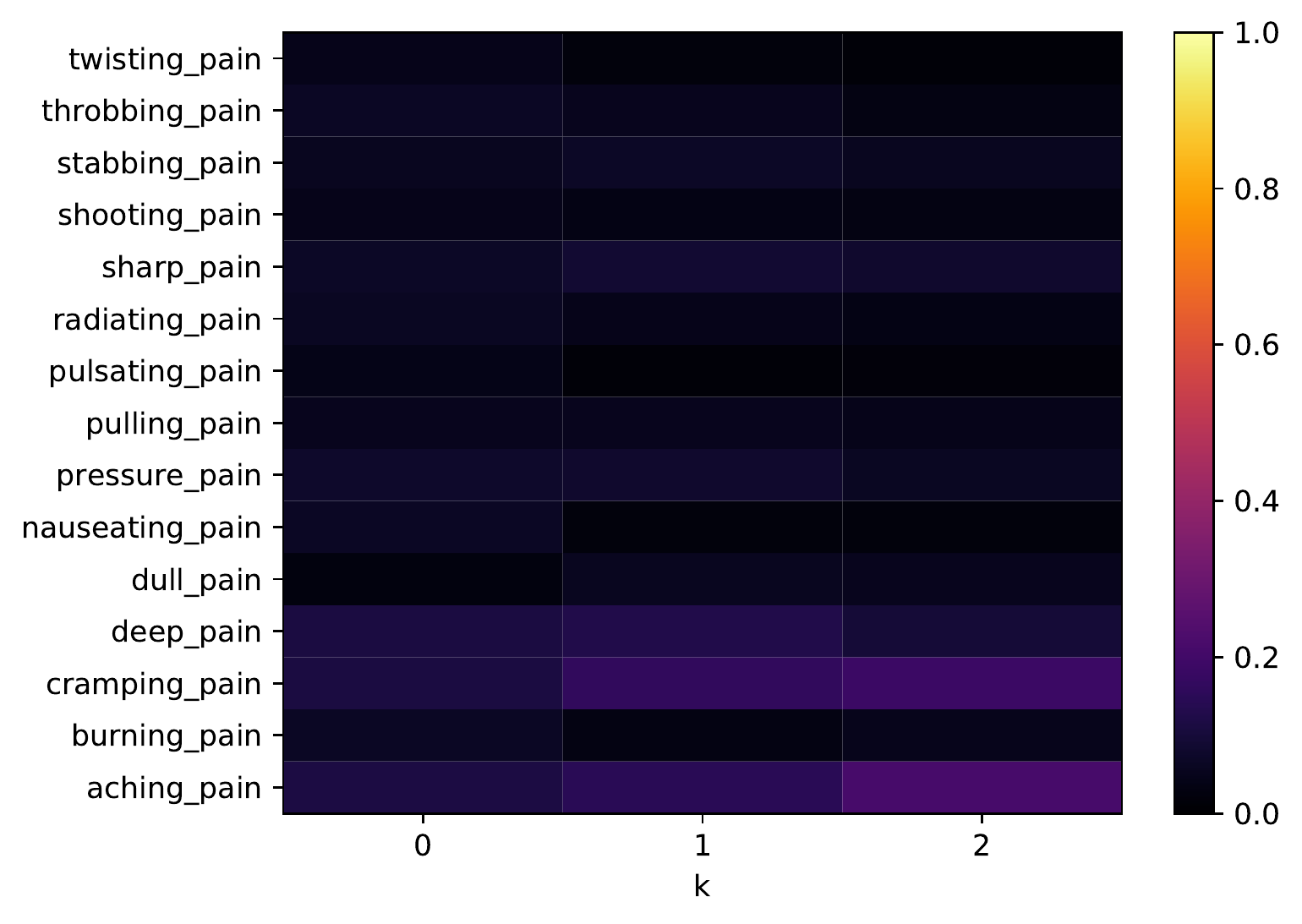}
		\caption{Describe the pain.}
	\end{subfigure}
	\begin{subfigure}[h]{0.30\textwidth}
		\includegraphics[width=\textwidth]{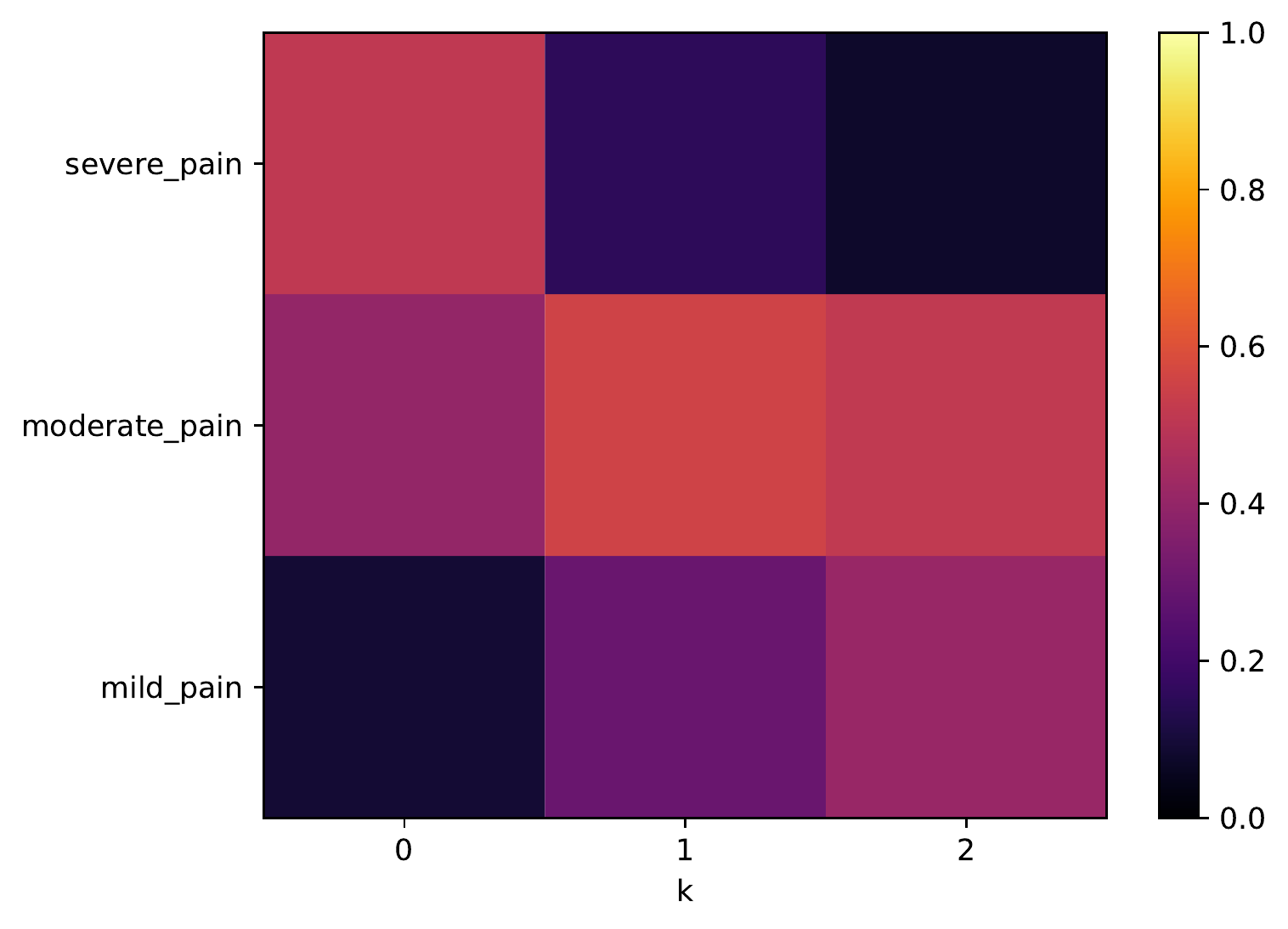}
		\caption{How severe is the pain?}
	\end{subfigure}
	
	\begin{subfigure}[h]{0.34\textwidth}
		\includegraphics[width=\textwidth]{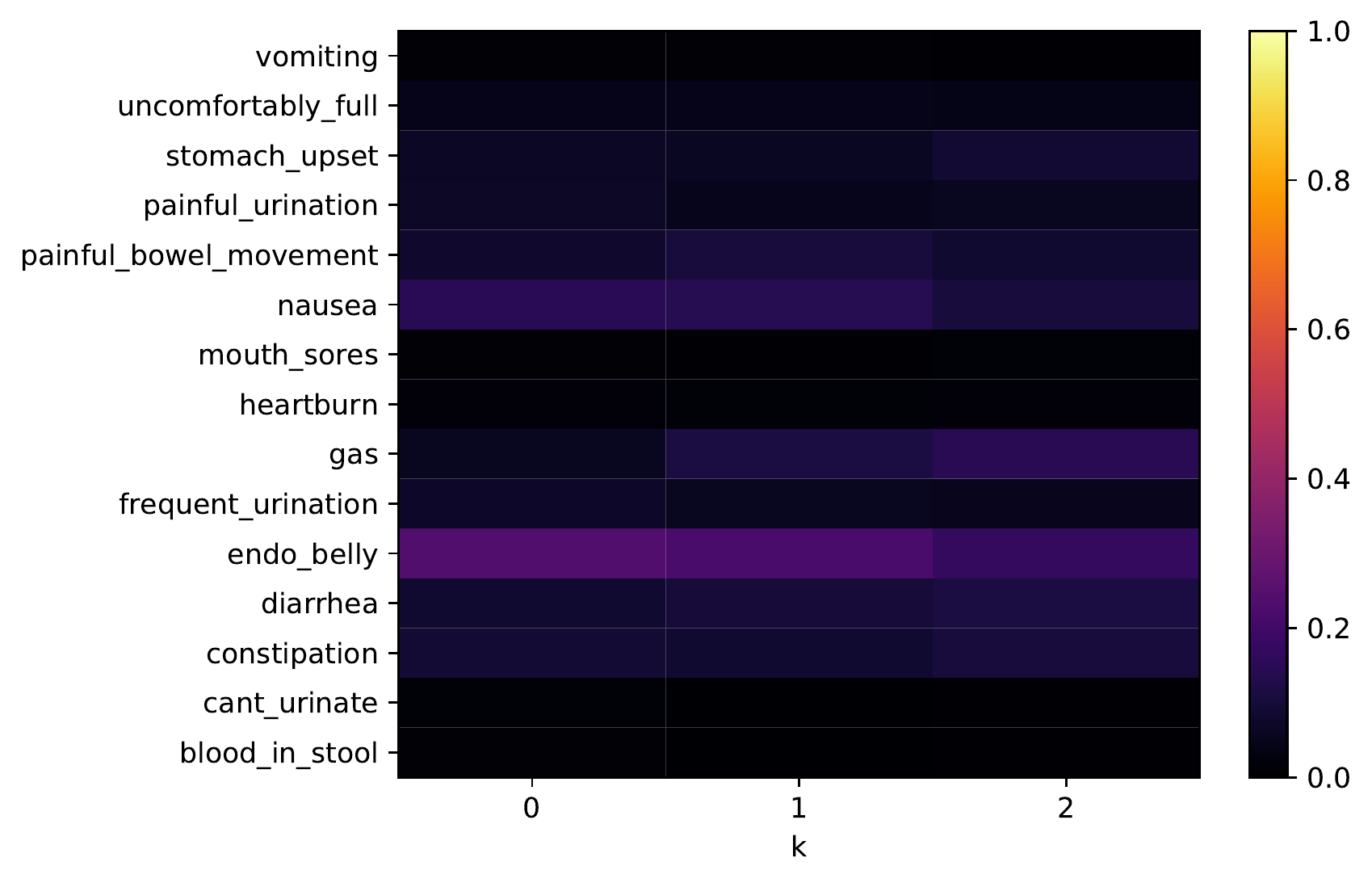}
		\caption{Describe GI/GU system.}
	\end{subfigure}
	\begin{subfigure}[h]{0.3\textwidth}
		\includegraphics[width=\textwidth]{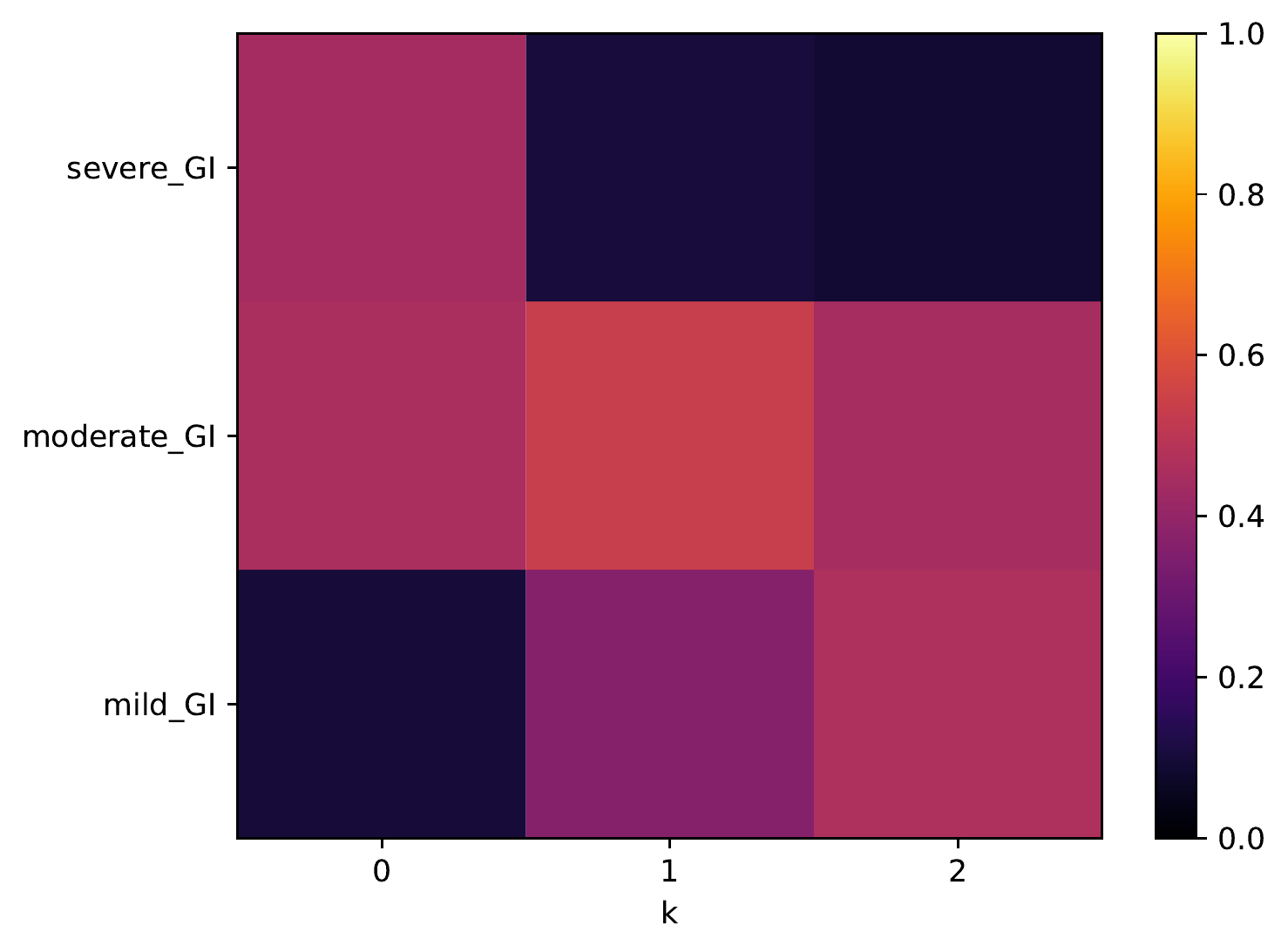}
		\caption{How severe is it.}
	\end{subfigure}
	\begin{subfigure}[h]{0.3\textwidth}
		\includegraphics[width=\textwidth]{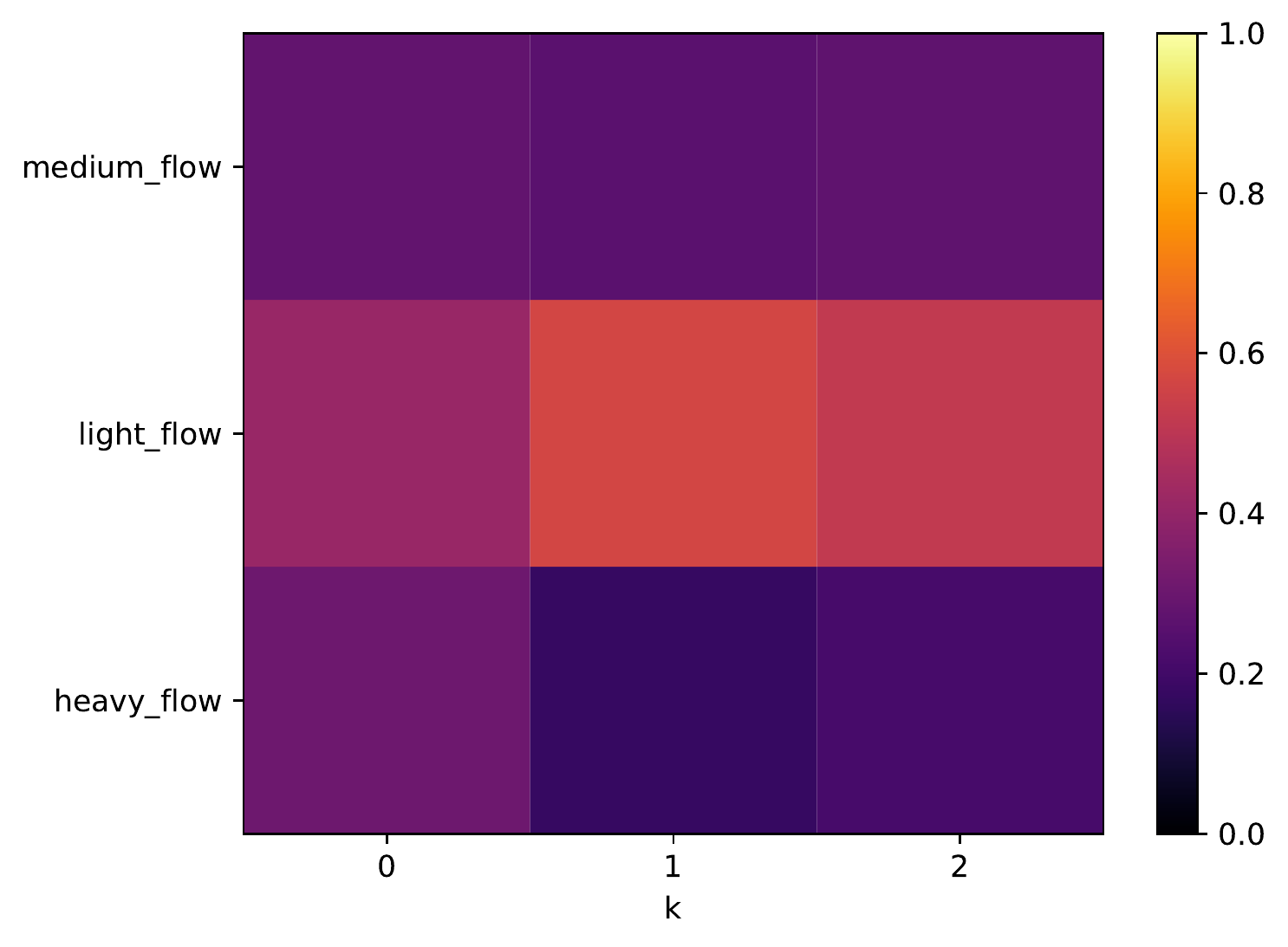}
		\caption{Describe the flow.}
	\end{subfigure}
	
	\begin{subfigure}[h]{0.31\textwidth}
		\includegraphics[width=\textwidth]{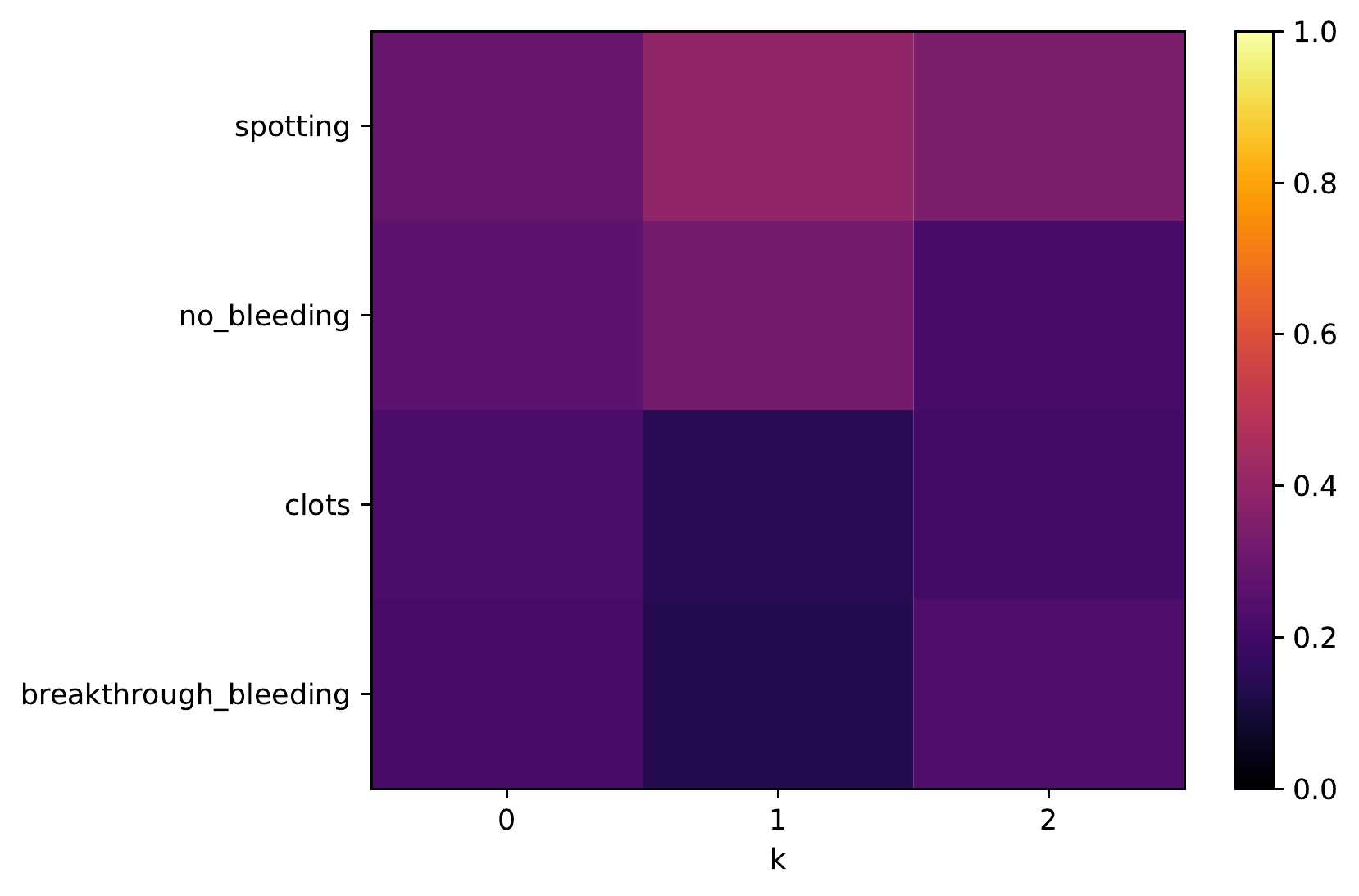}
		\caption{What kind of bleeding.}
	\end{subfigure}
	\begin{subfigure}[h]{0.31\textwidth}
		\includegraphics[width=\textwidth]{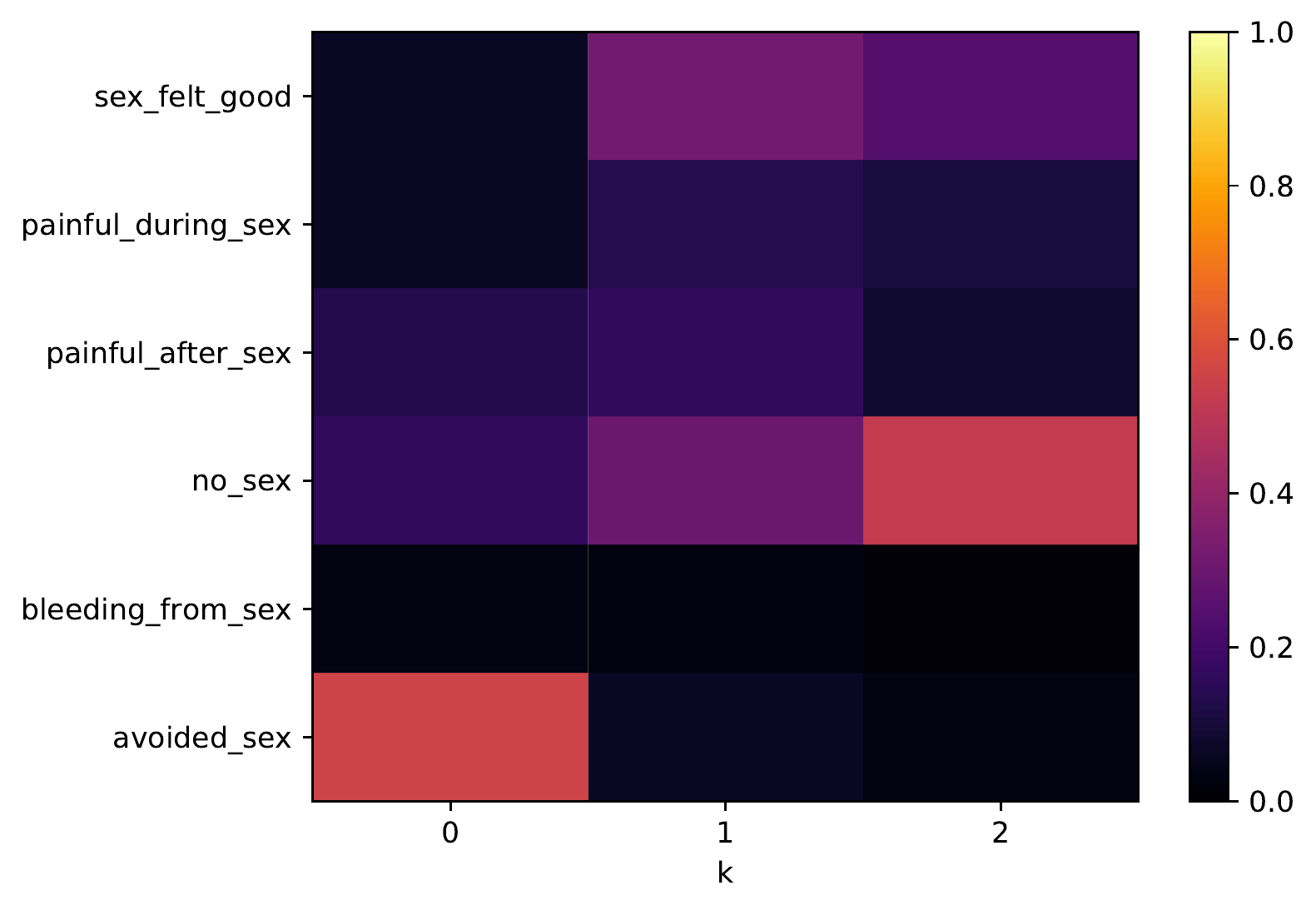}
		\caption{Describe sex.}
	\end{subfigure}
	\begin{subfigure}[h]{0.35\textwidth}
		\includegraphics[width=\textwidth]{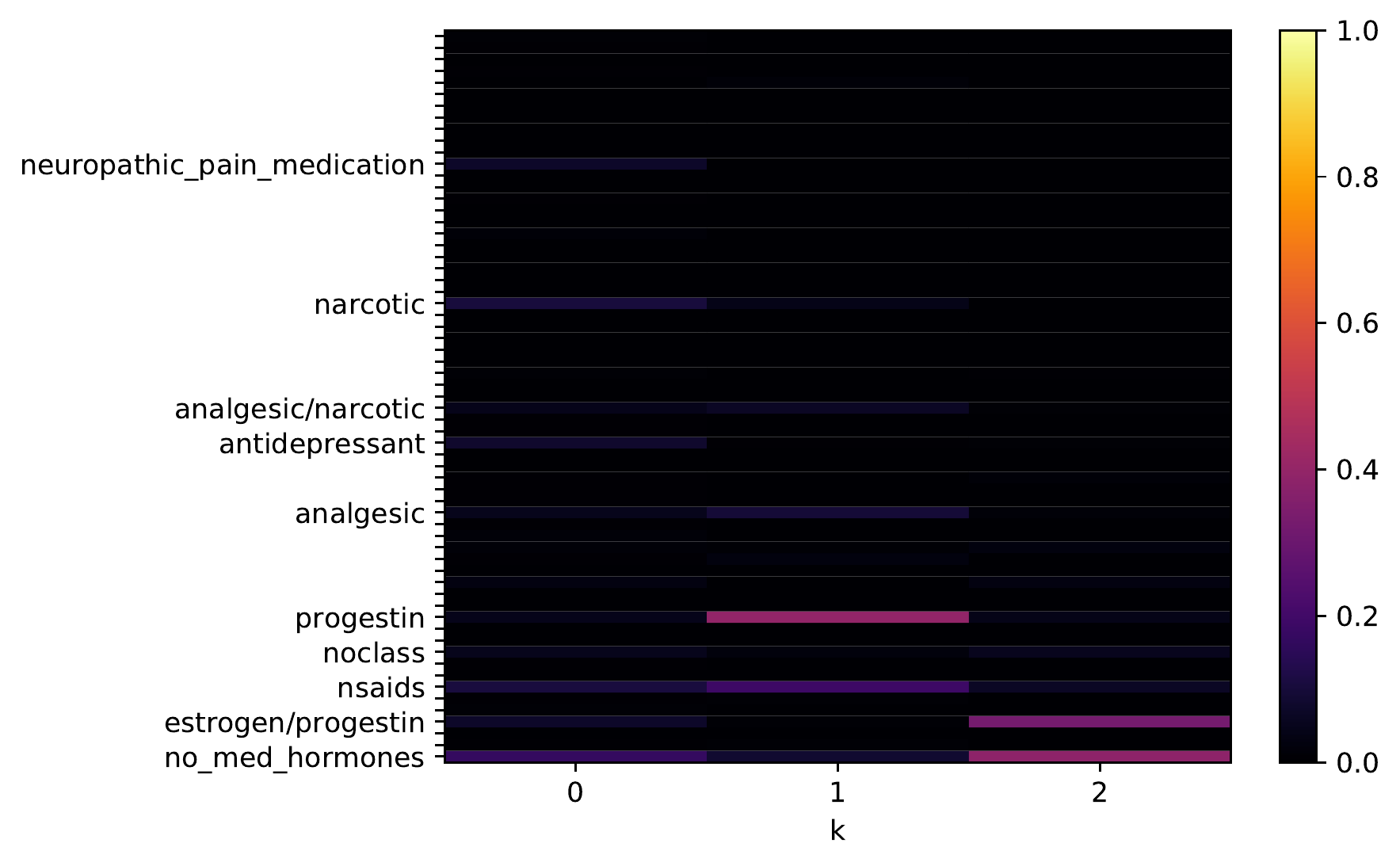}
		\caption{Medications/hormones taken.}
	\end{subfigure}
	
	\begin{subfigure}[h]{0.31\textwidth}
		\includegraphics[width=\textwidth]{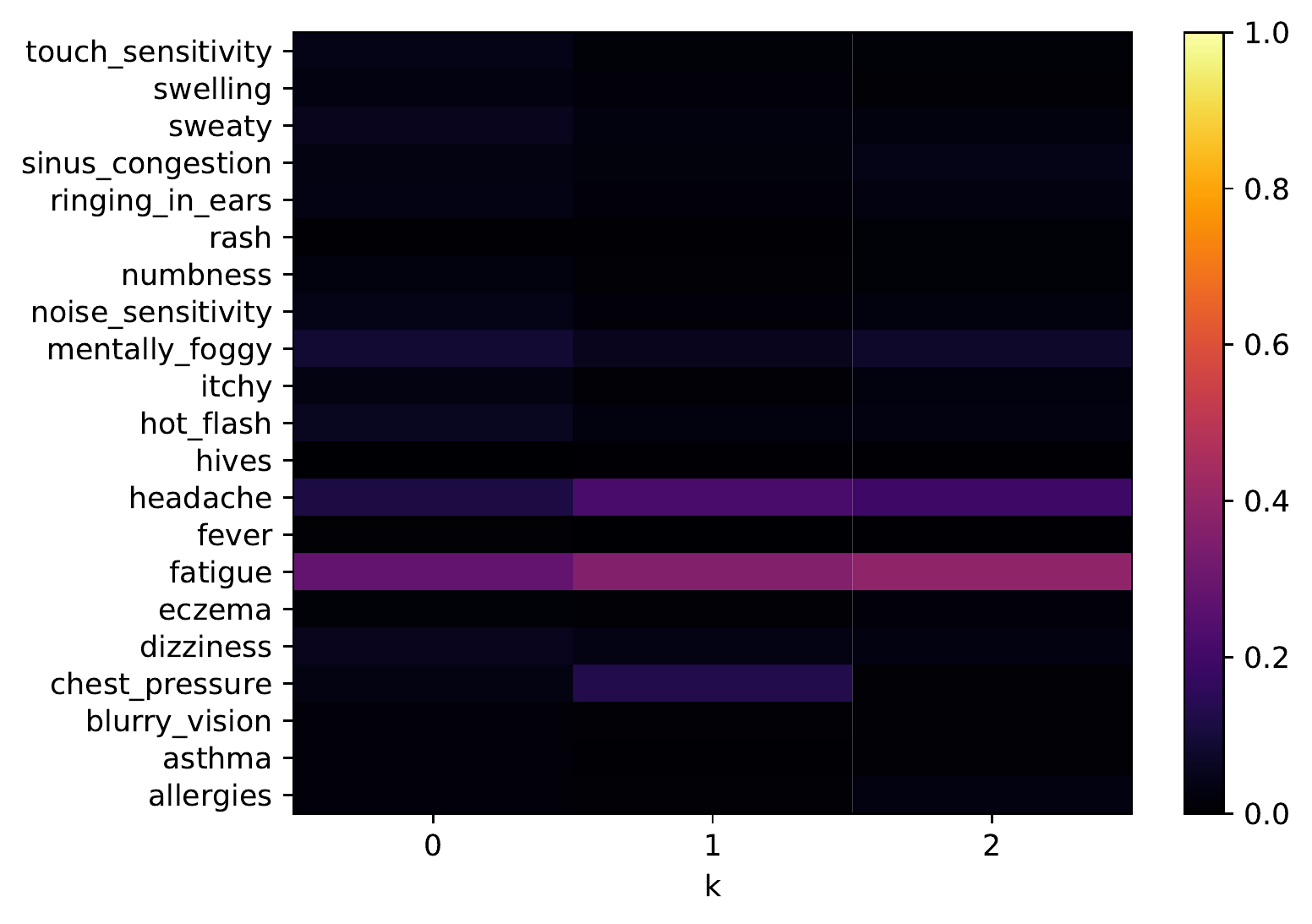}
		\caption{What are you experiencing.}
	\end{subfigure}
	\begin{subfigure}[h]{0.31\textwidth}
		\includegraphics[width=\textwidth]{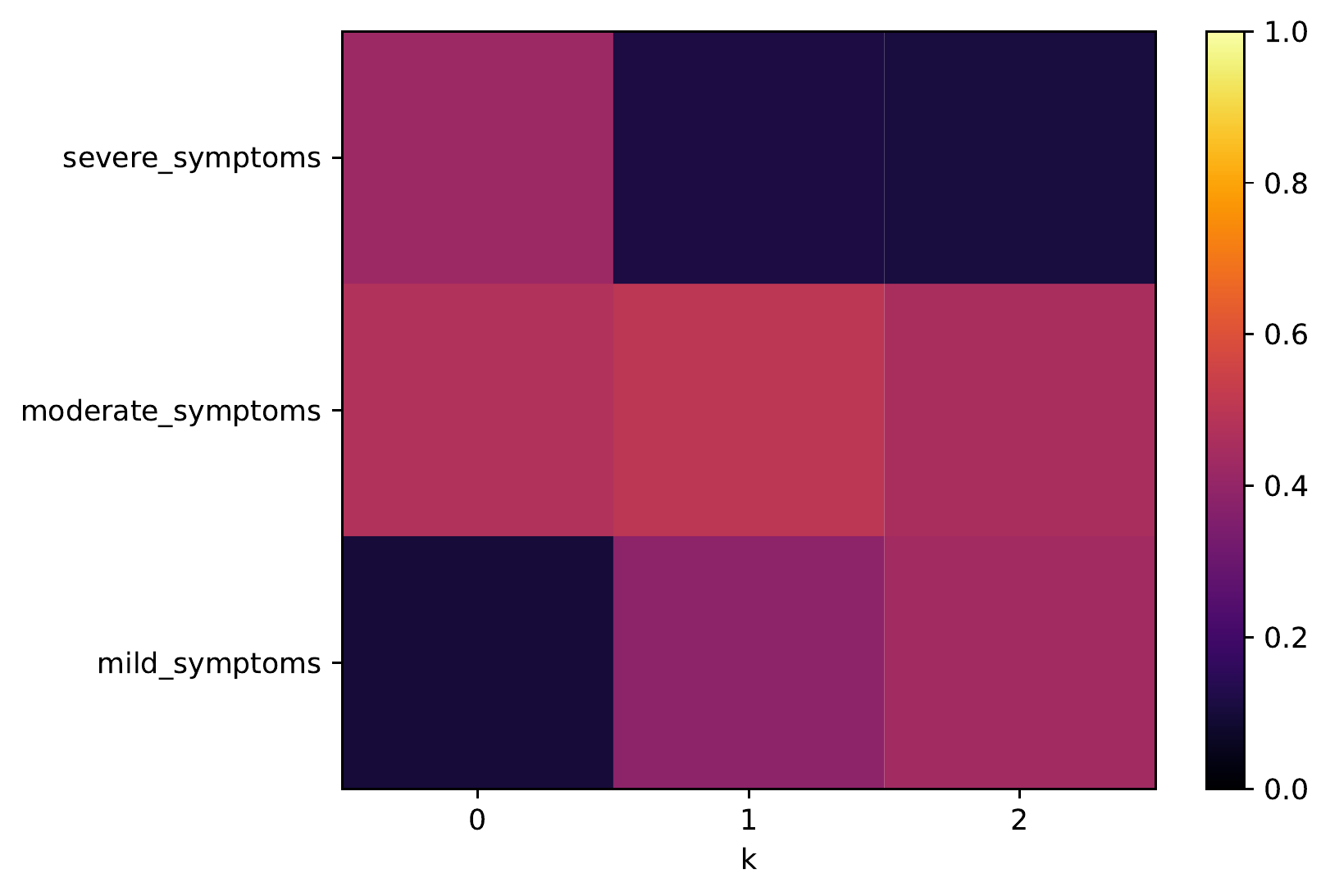}
		\caption{How severe is the symptom.}
	\end{subfigure}
	
	\begin{subfigure}[h]{0.31\textwidth}
		\includegraphics[width=\textwidth]{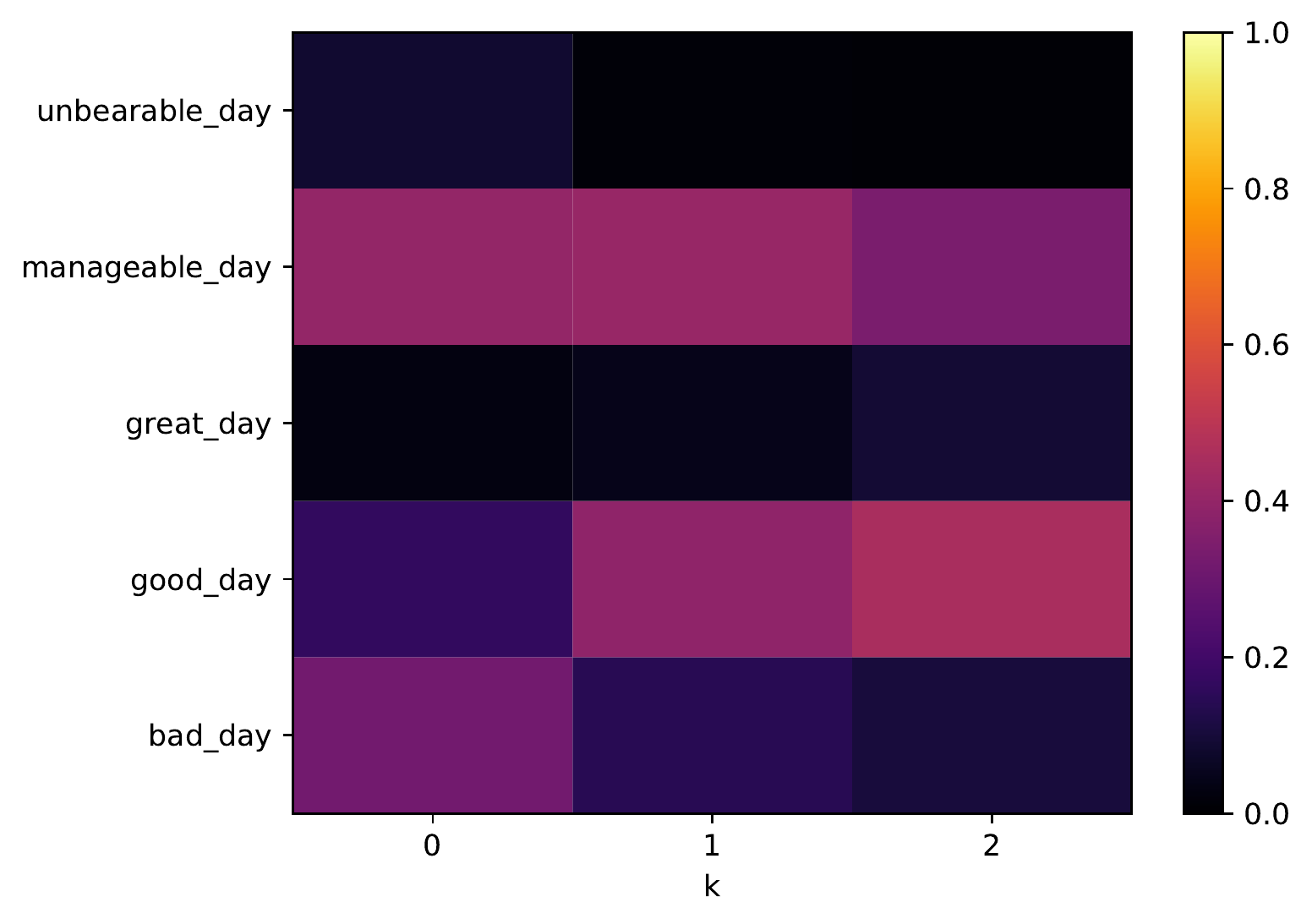}
		\caption{How was your day?}
	\end{subfigure}
	\begin{subfigure}[h]{0.31\textwidth}
		\includegraphics[width=\textwidth]{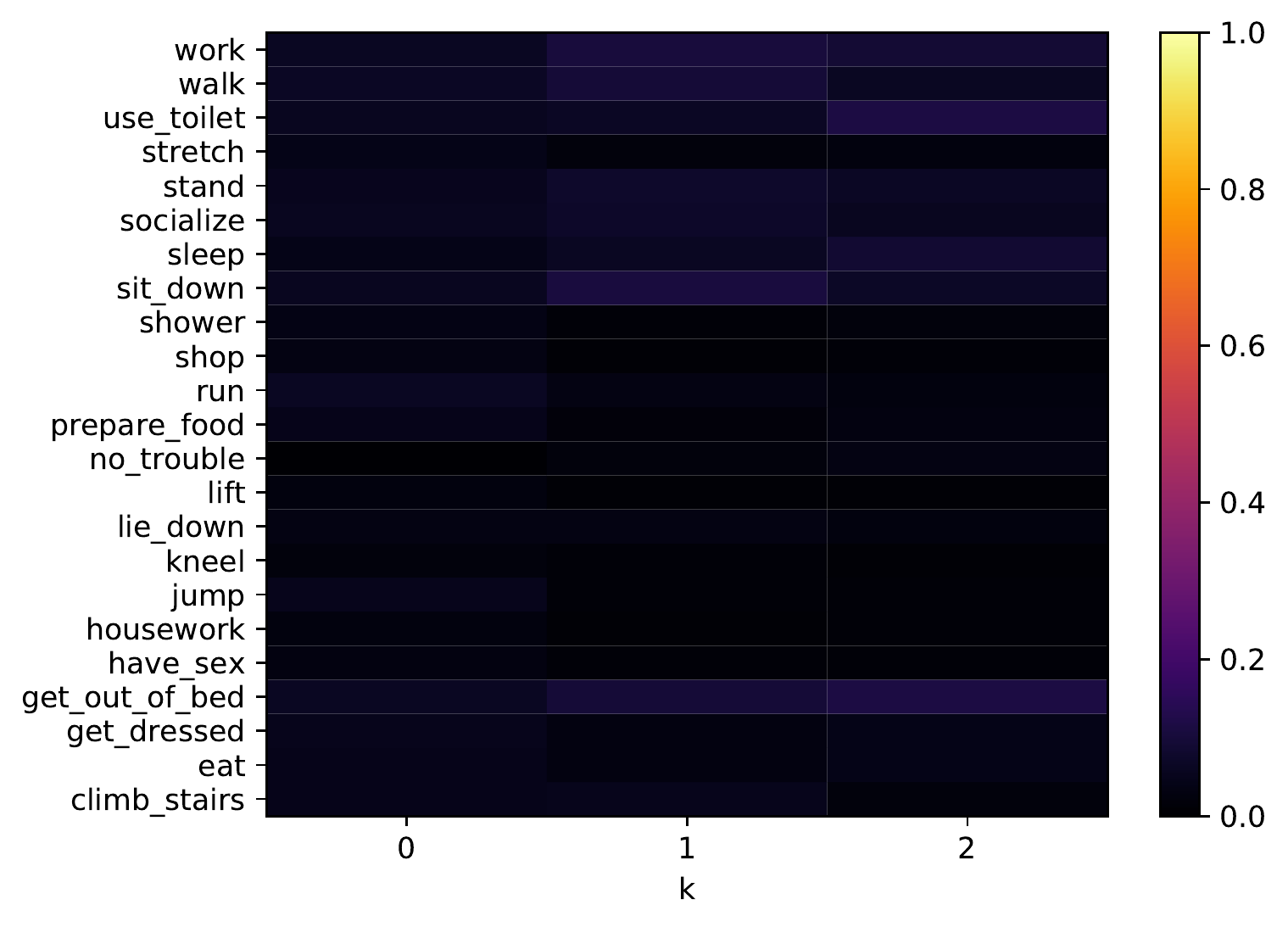}
		\caption{Activities}
	\end{subfigure}
	\caption{Visualization of per-question posteriors for learned endometriosis phenotypes.}
	\label{fig:posteriors}
\end{figure*}

\paragraph*{Model Likelihood.} 
We study the proposed model's phenotyping accuracy via 5-fold
log-likelihood evaluation. As shown in Figure \ref{fig:test_logliks}, there is a
significant improvement of our method when compared to vanilla LDA, where
responses to all questions are modeled together as a bag-of-words. As we allow
for per-question modalities, our method is capable of capturing discriminative signals in each of the variables, thus distinguishing between phenotypes (Figure~\ref{fig:posteriors}). We emphasize the robustness of the inference mechanism with respect to particular choices of hyperparameters. There is a slight performance improvement for sparse phenotypes, which we further take advantage of for interpretability purposes. 

\paragraph*{Interpretability.} 
In this experiment and those to follow, we focus on a selected model with 3 phenotypes 
(as models with more subtypes did not capture new discriminating insights) and sparse parameters ($\alpha=\beta=0.001$). The sparsity of the model allows for (i) few vocabulary items per-question being distinctive for each phenotype, and (ii) having discriminative per-participant phenotypic profiles (\ie located on the vertices of the probability simplex and thus strongly clustered).

\begin{figure}[!h]
	\centering
	\includegraphics[width=0.7\textwidth]{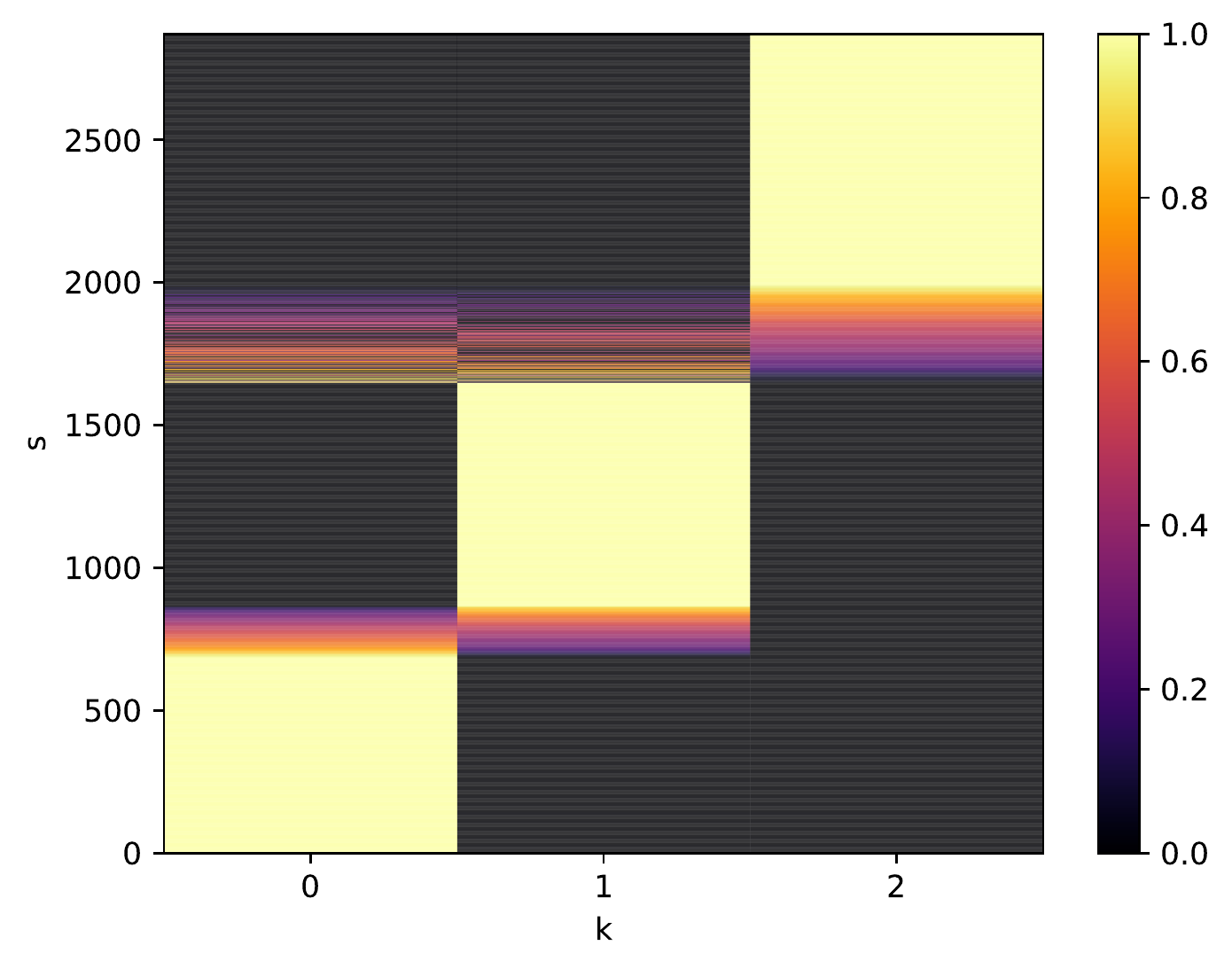}
	\caption{Learned per-participant posterior phenotypic distribution.}    
	\label{fig:participant_to_phenotype}
\end{figure}

The proposed model is clinically useful in that, for representing a participant, it allows for the tracked symptoms to be explained by a mixture of the learned phenotypes. We show the phenotypic assignments of participants to each learned endometriosis subtype in Figure \ref{fig:participant_to_phenotype}, and note that they do not correlate with the number of days (or observations) participants tracked. Although participants in all phenotypes have tracked similar number of days (39, 43 and 46 on average), participants associated with phenotype 0 have tracked more observations (on average, 126, 80 and 80, respectively). 

\begin{figure*}[!h]
        \centering
        \begin{subfigure}[h]{0.31\textwidth}
                \fbox{\includegraphics[width=\textwidth]{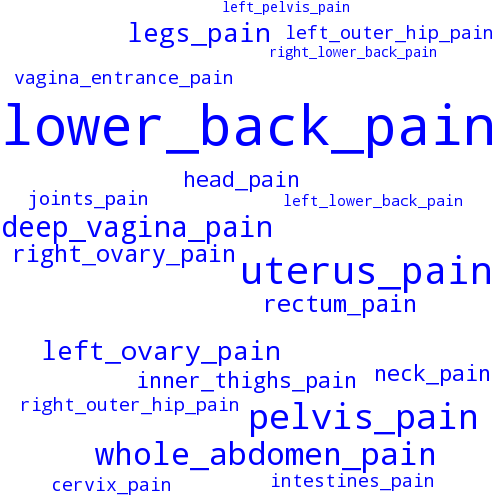}}
                \caption{Phenotype $k=0$}
        \end{subfigure}
        \begin{subfigure}[h]{0.31\textwidth}
                \fbox{\includegraphics[width=\textwidth]{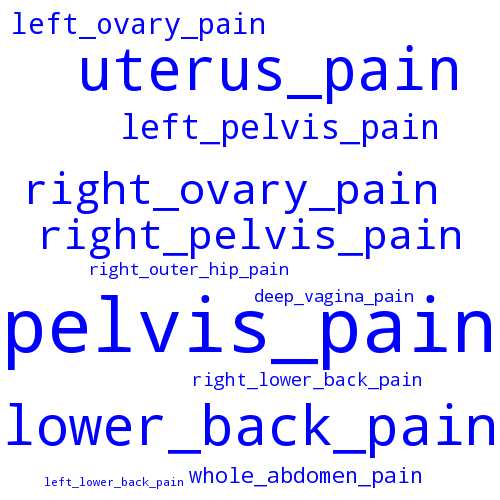}}
                \caption{Phenotype $k=1$}
        \end{subfigure}
        \begin{subfigure}[h]{0.31\textwidth}
                \fbox{\includegraphics[width=\textwidth]{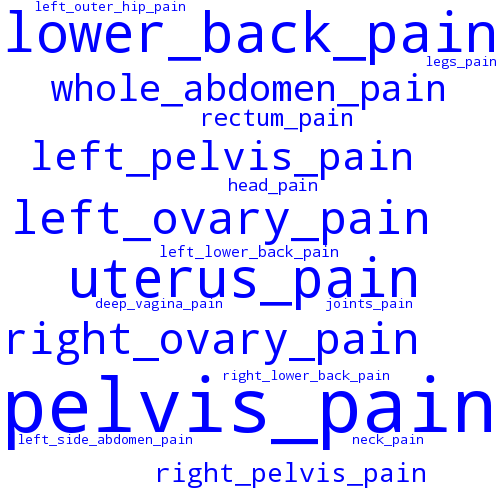}}
                \caption{Phenotype $k=2$}
        \end{subfigure}
        \caption{Where is the pain.}    
        \label{fig:wc_pain_where}
\end{figure*}

\begin{figure*}[!h]
        \centering
        \begin{subfigure}[h]{0.31\textwidth}
                \fbox{\includegraphics[width=\textwidth]{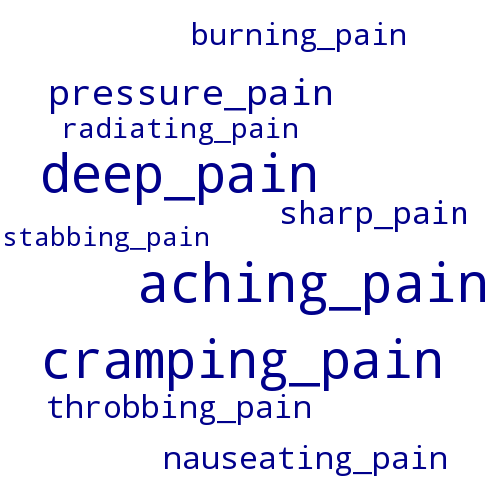}}
                \caption{Phenotype $k=0$}
        \end{subfigure}
        \begin{subfigure}[h]{0.31\textwidth}
                \fbox{\includegraphics[width=\textwidth]{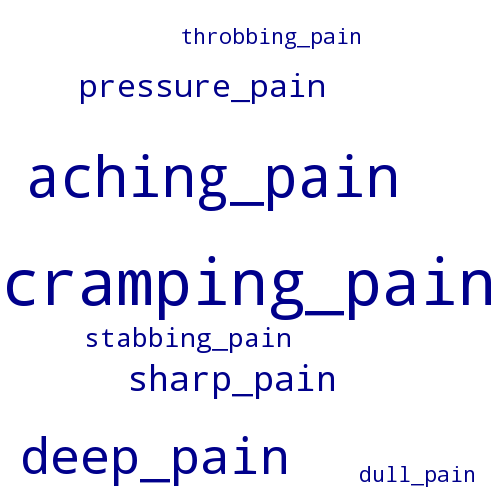}}
                \caption{Phenotype $k=1$}
        \end{subfigure}
        \begin{subfigure}[h]{0.31\textwidth}
                \fbox{\includegraphics[width=\textwidth]{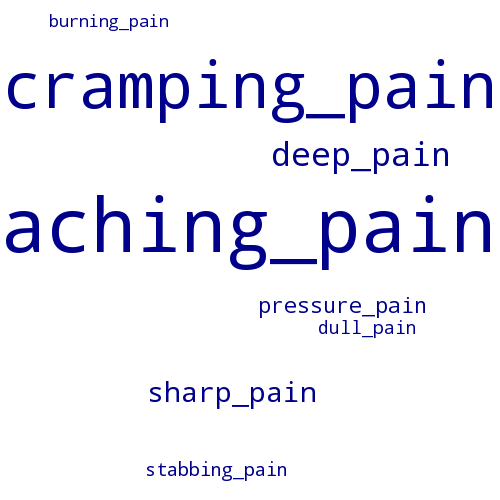}}
                \caption{Phenotype $k=2$}
        \end{subfigure}
        \caption{Describe the pain.}    
        \label{fig:wc_pain_describe}
\end{figure*}

\begin{figure*}[!h]
        \centering
        \begin{subfigure}[h]{0.31\textwidth}
                \fbox{\includegraphics[width=\textwidth]{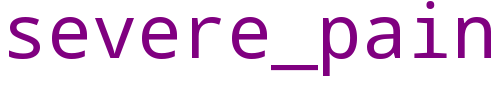}}
                \caption{Phenotype $k=0$}
        \end{subfigure}
        \begin{subfigure}[h]{0.31\textwidth}
                \fbox{\includegraphics[width=\textwidth]{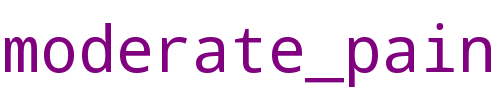}}
                \caption{Phenotype $k=1$}
        \end{subfigure}
        \begin{subfigure}[h]{0.31\textwidth}
                \fbox{\includegraphics[width=\textwidth]{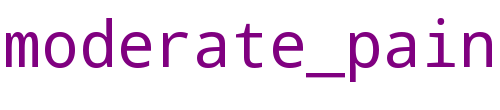}}
                \caption{Phenotype $k=2$}
        \end{subfigure}
        \caption{How severe is the pain.}
        \label{fig:wc_pain_severity}
\end{figure*}

\begin{figure*}[!h]
        \centering
        \begin{subfigure}[h]{0.31\textwidth}
                \fbox{\includegraphics[width=\textwidth]{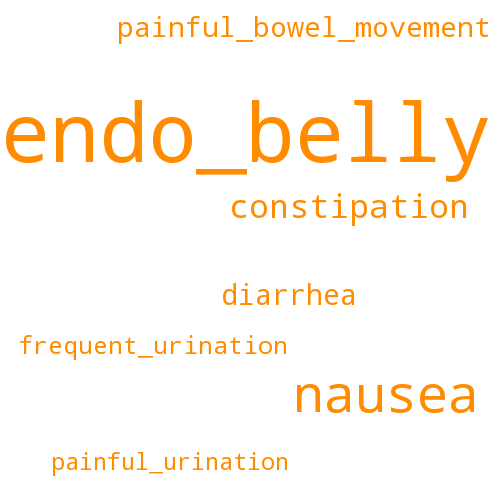}}
                \caption{Phenotype $k=0$}
        \end{subfigure}
        \begin{subfigure}[h]{0.31\textwidth}
                \fbox{\includegraphics[width=\textwidth]{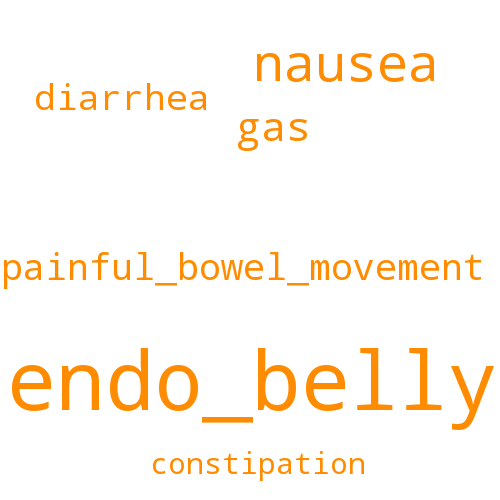}}
                \caption{Phenotype $k=1$}
        \end{subfigure}
        \begin{subfigure}[h]{0.31\textwidth}
                \fbox{\includegraphics[width=\textwidth]{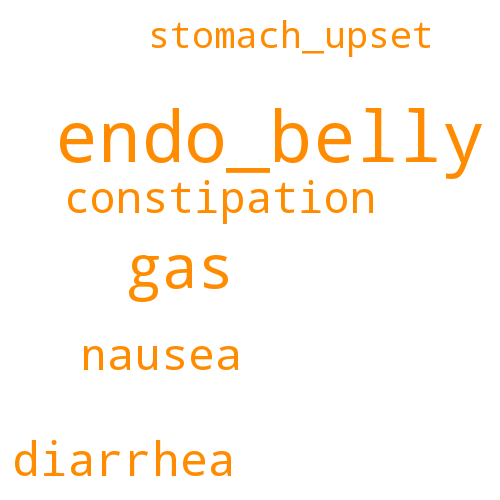}}
                \caption{Phenotype $k=2$}
        \end{subfigure}
        \caption{Describe GI/GU system.}
        \label{fig:wc_gigu_describe}
\end{figure*}

\begin{figure*}[!h]
        \centering
        \begin{subfigure}[h]{0.31\textwidth}
                \fbox{\includegraphics[width=\textwidth]{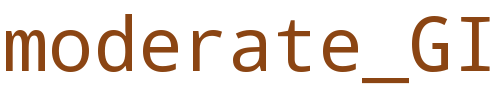}}
                \caption{Phenotype $k=0$}
        \end{subfigure}
        \begin{subfigure}[h]{0.31\textwidth}
                \fbox{\includegraphics[width=\textwidth]{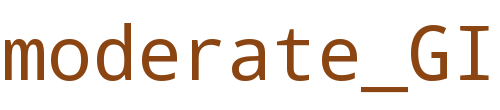}}
                \caption{Phenotype $k=1$}
        \end{subfigure}
        \begin{subfigure}[h]{0.31\textwidth}
                \fbox{\includegraphics[width=\textwidth]{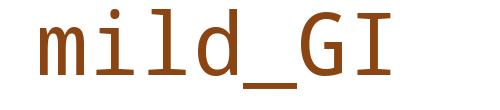}}
                \caption{Phenotype $k=2$}
        \end{subfigure}
        \caption{How severe is it.}
        \label{fig:wc_gigu_severity}
\end{figure*}

\begin{figure*}[!h]
        \centering
        \begin{subfigure}[h]{0.31\textwidth}
                \fbox{\includegraphics[width=\textwidth]{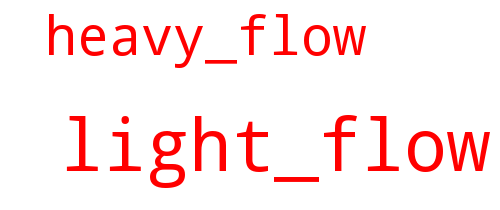}}
                \caption{Phenotype $k=0$}
        \end{subfigure}
        \begin{subfigure}[h]{0.31\textwidth}
                \fbox{\includegraphics[width=\textwidth]{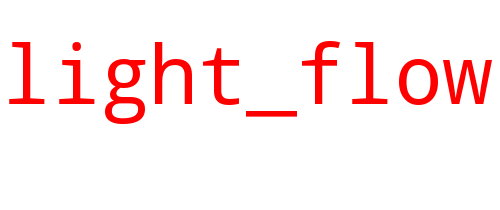}}
                \caption{Phenotype $k=1$}
        \end{subfigure}
        \begin{subfigure}[h]{0.31\textwidth}
                \fbox{\includegraphics[width=\textwidth]{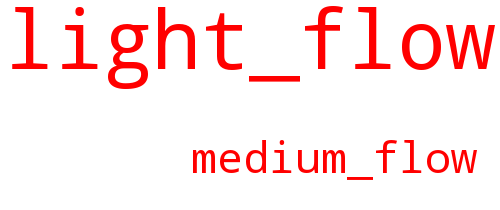}}
                \caption{Phenotype $k=2$}
        \end{subfigure}
        \caption{Describe the flow.}
        \label{fig:wc_flow}
\end{figure*}

\begin{figure*}[!h]
        \centering
        \begin{subfigure}[h]{0.31\textwidth}
                \fbox{\includegraphics[width=\textwidth]{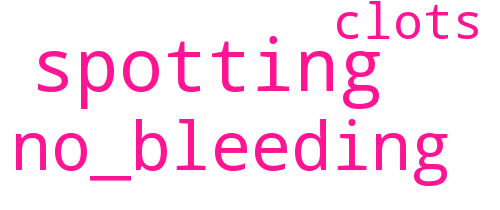}}
                \caption{Phenotype $k=0$}
        \end{subfigure}
        \begin{subfigure}[h]{0.31\textwidth}
                \fbox{\includegraphics[width=\textwidth]{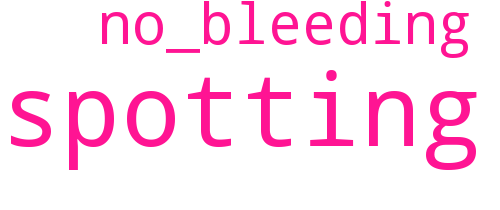}}
                \caption{Phenotype $k=1$}
        \end{subfigure}
        \begin{subfigure}[h]{0.31\textwidth}
                \fbox{\includegraphics[width=\textwidth]{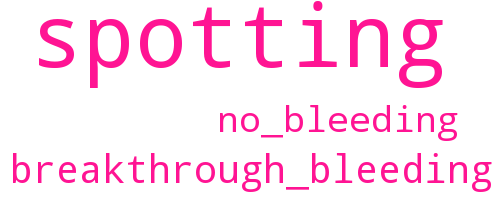}}
                \caption{Phenotype $k=2$}
        \end{subfigure}
        \caption{What kind of bleeding.}
        \label{fig:wc_bleeding}
\end{figure*}

\begin{figure*}[!h]
        \centering
        \begin{subfigure}[h]{0.31\textwidth}
                \fbox{\includegraphics[width=\textwidth]{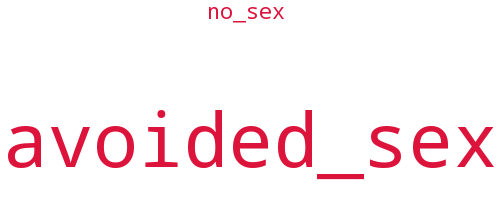}}
                \caption{Phenotype $k=0$}
        \end{subfigure}
        \begin{subfigure}[h]{0.31\textwidth}
                \fbox{\includegraphics[width=\textwidth]{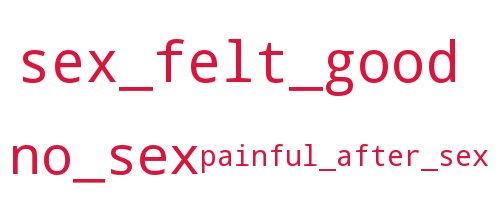}}
                \caption{Phenotype $k=1$}
        \end{subfigure}
        \begin{subfigure}[h]{0.31\textwidth}
                \fbox{\includegraphics[width=\textwidth]{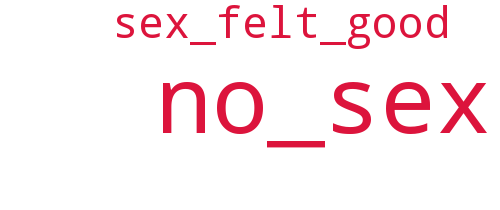}}
                \caption{Phenotype $k=2$}
        \end{subfigure}
        \caption{Describe sex.}
        \label{fig:wc_sex}
\end{figure*}

\begin{figure*}[!h]
        \centering
        \begin{subfigure}[h]{0.31\textwidth}
                \fbox{\includegraphics[width=\textwidth]{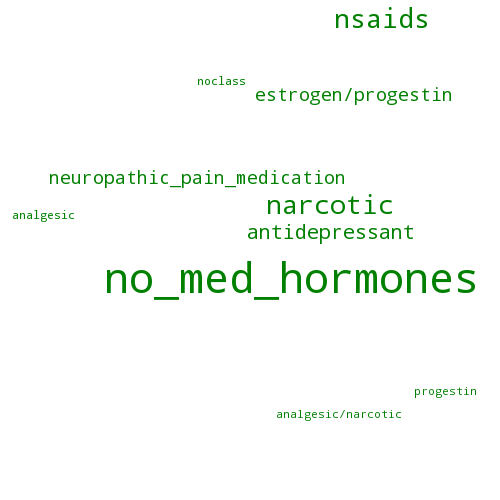}}
                \caption{Phenotype $k=0$}
        \end{subfigure}
        \begin{subfigure}[h]{0.31\textwidth}
                \fbox{\includegraphics[width=\textwidth]{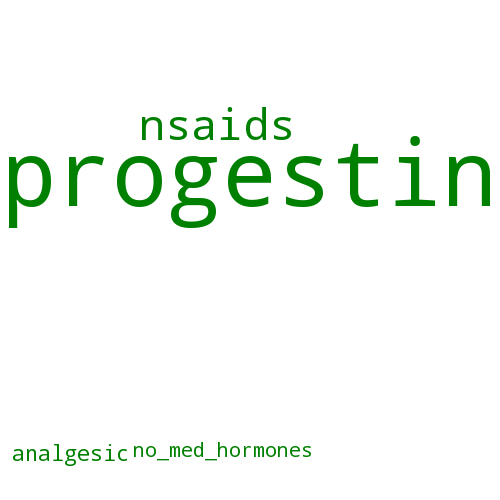}}
                \caption{Phenotype $k=1$}
        \end{subfigure}
        \begin{subfigure}[h]{0.31\textwidth}
                \fbox{\includegraphics[width=\textwidth]{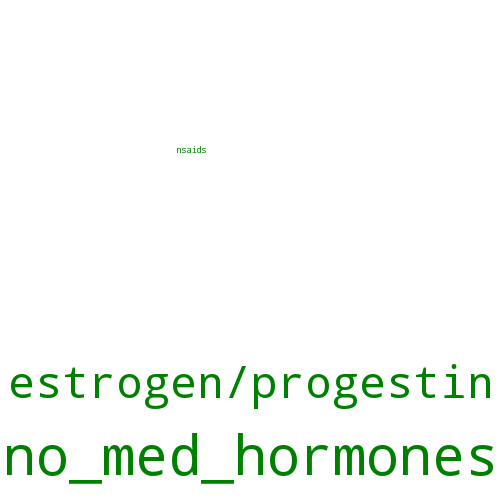}}
                \caption{Phenotype $k=2$}
        \end{subfigure}
        \caption{Medications/hormones taken.}
        \label{fig:wc_meds_hormones}
\end{figure*}

\begin{figure*}[!h]
        \centering
        \begin{subfigure}[h]{0.31\textwidth}
                \fbox{\includegraphics[width=\textwidth]{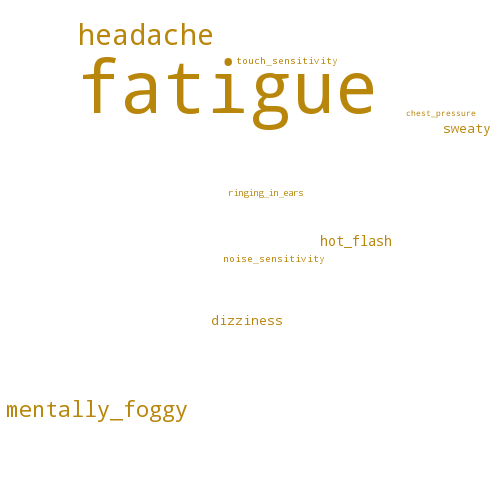}}
                \caption{Phenotype $k=0$}
        \end{subfigure}
        \begin{subfigure}[h]{0.31\textwidth}
                \fbox{\includegraphics[width=\textwidth]{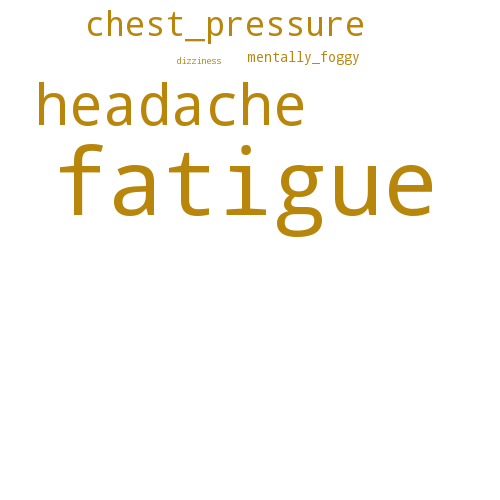}}
                \caption{Phenotype $k=1$}
        \end{subfigure}
        \begin{subfigure}[h]{0.31\textwidth}
                \fbox{\includegraphics[width=\textwidth]{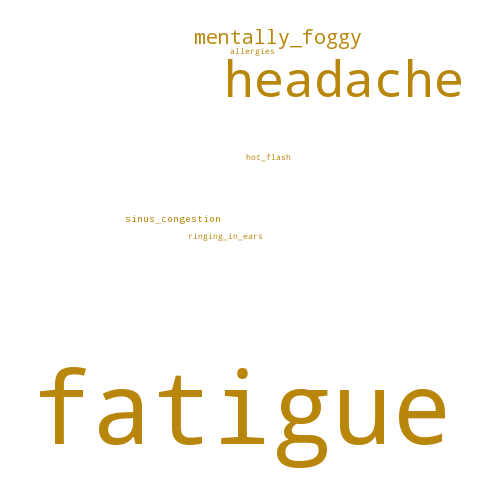}}
                \caption{Phenotype $k=2$}
        \end{subfigure}
        \caption{What are you experiencing.}
        \label{fig:wc_symptoms_describe}
\end{figure*}

\begin{figure*}[!h]
        \centering
        \begin{subfigure}[h]{0.31\textwidth}
                \fbox{\includegraphics[width=\textwidth]{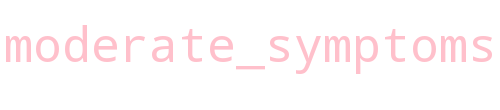}}
                \caption{Phenotype $k=0$}
        \end{subfigure}
        \begin{subfigure}[h]{0.31\textwidth}
                \fbox{\includegraphics[width=\textwidth]{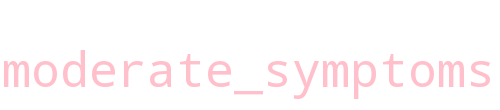}}
                \caption{Phenotype $k=1$}
        \end{subfigure}
        \begin{subfigure}[h]{0.31\textwidth}
                \fbox{\includegraphics[width=\textwidth]{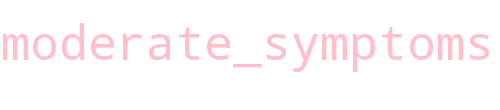}}
                \caption{Phenotype $k=2$}
        \end{subfigure}
        \caption{How severe is the symptom.}
        \label{fig:wc_symptoms_severity}
\end{figure*}

\begin{figure*}[!h]
        \centering
        \begin{subfigure}[h]{0.31\textwidth}
                \fbox{\includegraphics[width=\textwidth]{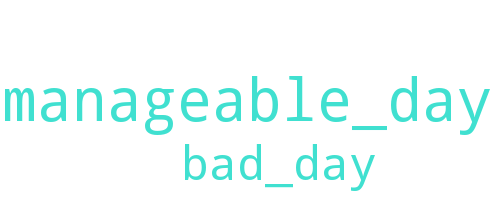}}
                \caption{Phenotype $k=0$}
        \end{subfigure}
        \begin{subfigure}[h]{0.31\textwidth}
                \fbox{\includegraphics[width=\textwidth]{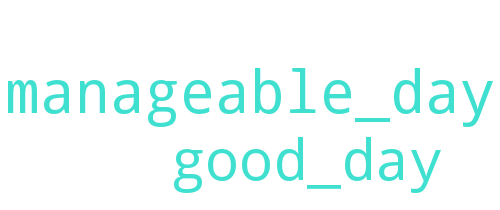}}
                \caption{Phenotype $k=1$}
        \end{subfigure}
        \begin{subfigure}[h]{0.31\textwidth}
                \fbox{\includegraphics[width=\textwidth]{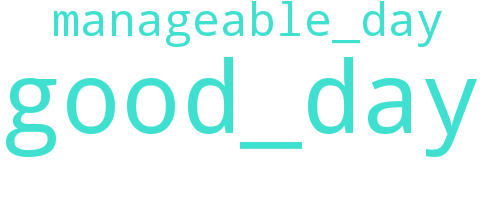}}
                \caption{Phenotype $k=2$}
        \end{subfigure}
        \caption{How was your day?}
        \label{fig:wc_day}
\end{figure*}

\begin{figure*}[!h]
        \centering
        \begin{subfigure}[h]{0.31\textwidth}
                \fbox{\includegraphics[width=\textwidth]{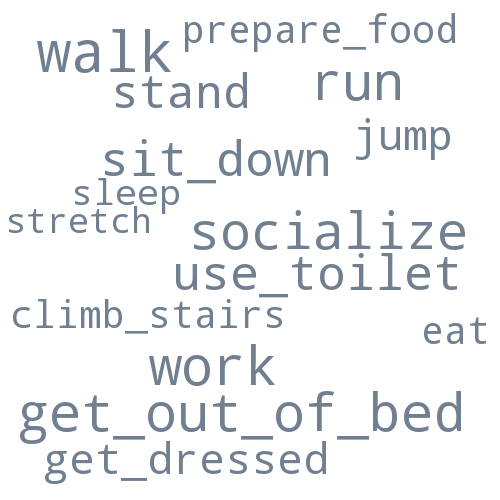}}
                \caption{Phenotype $k=0$}
        \end{subfigure}
        \begin{subfigure}[h]{0.31\textwidth}
                \fbox{\includegraphics[width=\textwidth]{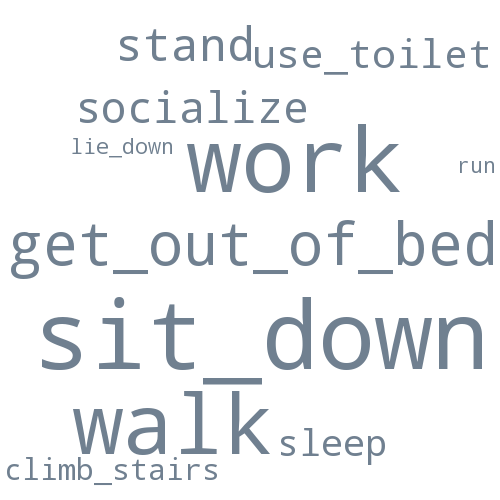}}
                \caption{Phenotype $k=1$}
        \end{subfigure}
        \begin{subfigure}[h]{0.31\textwidth}
                \fbox{\includegraphics[width=\textwidth]{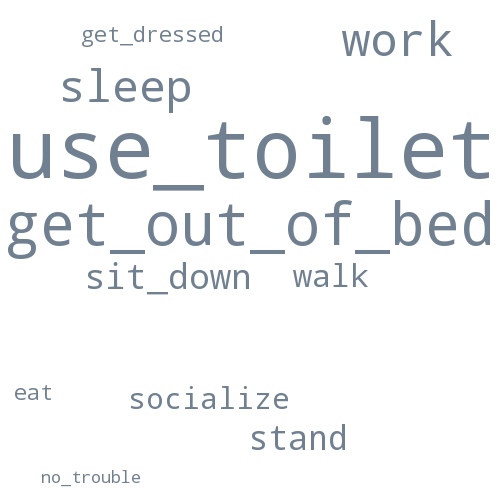}}
                \caption{Phenotype $k=2$}
        \end{subfigure}
        \caption{Activities}
        \label{fig:wc_activities}
\end{figure*}

We now elaborate on the specificity of signs and symptoms of the disease for each learned phenotype, based on learned per-question posteriors (Figure~\ref{fig:posteriors}), and their word-cloud visualizations (Figures~\ref{fig:wc_pain_where} -- \ref{fig:wc_activities}).

Pain is shared across all phenotypes, and its location is most often tracked for pelvic and lower back areas (Figure~\ref{fig:wc_pain_where}).
The commonality of these pain locations is expected, given that lower back and pelvic pain are stereotypical for endometriosis \citep{Chiantera2017}. Phenotype 0,
however, has a much wider range of locations in which pain is experienced, characterized by reports of deep vagina pain, rectal pain, and pain projecting along
the legs (Figure~\ref{fig:wc_pain_where}). Similarly, certain descriptions of pain are common across phenotypes, such as deep, aching, and cramping (Figure~\ref{fig:wc_pain_describe}). However, phenotype 0 has a wider range of descriptions of how pain is experienced, besides pain being predominantly characterized as severe (Figure~\ref{fig:wc_pain_severity}).
These characterizations of pain across phenotypes indicate that the burden on pain is much heavier and severe for phenotype 0, while moderate or mild
descriptors are associated with phenotypes 1 and 2 (this trend is consistent for the majority of questions).

Each phenotype had strong involvement of gastrointestinal symptoms and was
characterized by ``endo belly" across phenotypes (Figure~\ref{fig:wc_gigu_describe}). This is a very common disease symptom where the abdomen severely bloats \citep{ek2015, luscombe2009}. Phenotype 0 is distinctively associated with genitourinary symptoms, like painful and frequent urination or dysuria. While genitourinary symptoms in endometriosis are known, their association with a subgroup of patients is novel \citep{denny2007}.   

As for menstrual characteristics (visualized in Figure~\ref{fig:wc_flow}), phenotype 0 reports heavier flow or menorrhagia. Menorrhagia is a common endometriosis symptom but has not been associated with a particular subgroup of endometriosis patients \citep{vercellini1997}.
Moreover, although all phenotypes have tracked spotting or bleeding outside of the period (Figure~\ref{fig:wc_bleeding}), phenotype 0 is also likely to track clots,
consistent with descriptions of menorrhagia \citep{warner2004}.

Painful sex or dyspareunia is a wieldy known symptom for endometriosis \citep{denny2007dyspareunia}. As shown in Figure~\ref{fig:wc_sex}, our model learned that all phenotypes avoided sex. Furthermore, phenotype 0 is distinguished by the active avoidance of sex, and does not experience satisfying sex.   

General quality of life, as measured by the ``How was your day?" question and tracked daily activities visualized in Figures~\ref{fig:wc_day} and \ref{fig:wc_activities} respectively, show similar severe, moderate, and mild patterns for each phenotype.  All phenotypes have manageable days, though bad days are more present for phenotype 0. Even if there are common difficulties associated with activities of daily living across phenotypes, phenotype 0 has a wider range of tracked problems.   

Medications and hormones clearly discriminate the patients by severity of disease, as shown in Figure~\ref{fig:wc_meds_hormones}. Phenotype 2 does not take medications, or may take a combination of hormonal medications like birth control pills (BCPs). BCPs are often used as a first line treatment for endometriosis symptoms \citep{schrager2013}. Phenotype 1 takes BCPs as well, but is further characterized by its use of analgesics. In stark contrast, phenotype 0 has heavy medication use, taking anti-depressants and strong pain medications,including narcotics, neuropathic pain medications and opioids.

Other symptoms of endometriosis, which were collected via the ``What else are you experiencing?" question shown in Figure~\ref{fig:wc_symptoms_describe},
reflect the chronic nature of endometriosis.
Fatigue, mental fogginess, and headache occur across all phenotypes and are similar to other complex chronic conditions like chronic fatigue syndrome.  These symptoms are also characteristic of low grade inflammation \citep{holgate2011, louati2015}.   

\clearpage

\paragraph*{Clustering agreement.}
Table \ref{tab:expert_phenotype_confusion} shows the agreement between the learned model's posterior assignments and those determined by two different experts, with purity values of $0.6$ and $0.73$ for the first and second experts, respectively. The expert assignments to model's phenotype 0 (the severe case) were the most accurate, while deciding between the other two phenotypes was deemed to be harder. As a matter of fact, both experts raised several concerns about needing more data to make more informed decisions, as they had difficulties distinguishing between non-severe endometriosis cases.

\begin{table}[!h]
	\begin{center}
			\begin{tabular}{*{5}{|c}|}
				\hline
				\multicolumn{2}{|c|}{\multirow{2}{*}{Phenotype}} & \multicolumn{3}{c|}{Expert 1 \cellcolor[gray]{0.6}} \\ \cline{3-5}
				\multicolumn{1}{|c}{} & \multicolumn{1}{c|}{} & \cellcolor[gray]{0.8} 0 & \cellcolor[gray]{0.8} 1 & \cellcolor[gray]{0.8} 2 \\ \hline
				\cellcolor[gray]{0.6} & \cellcolor[gray]{0.8} 0 & 8 & 2 & 0 \\ \cline{1-5}
				\cellcolor[gray]{0.6} & \cellcolor[gray]{0.8} 1 & 3 & 5 & 2 \\ \cline{1-5}
				\cellcolor[gray]{0.6} & \cellcolor[gray]{0.8} 2 & 3 & 2 & 5 \\ \cline{1-5}
				\multicolumn{1}{c}{\multirow{-5}{*}{Model}} & \multicolumn{4}{c}{}
			\end{tabular}
			\begin{tabular}{*{5}{|c}|}
				\hline
				\multicolumn{2}{|c|}{\multirow{2}{*}{Phenotype}} & \multicolumn{3}{c|}{Expert 2\cellcolor[gray]{0.6}} \\ \cline{3-5}
				\multicolumn{1}{|c}{} & \multicolumn{1}{c|}{} & \cellcolor[gray]{0.8} 0 & \cellcolor[gray]{0.8} 1 & \cellcolor[gray]{0.8} 2 \\ \hline
				\cellcolor[gray]{0.6} & \cellcolor[gray]{0.8} 0 & 7 & 1 & 2 \\ \cline{1-5}
				\cellcolor[gray]{0.6} & \cellcolor[gray]{0.8} 1 & 1 & 8 & 1 \\ \cline{1-5}
				\cellcolor[gray]{0.6} & \cellcolor[gray]{0.8} 2 & 0 & 7 & 3 \\ \cline{1-5}
				\multicolumn{1}{c}{\multirow{-5}{*}{Model}} & \multicolumn{4}{c}{}
			\end{tabular}
		\vspace*{-0.5cm}
		\caption{Phenotype confusion matrices.}
		\label{tab:expert_phenotype_confusion}
	\end{center}
\end{table}
\vspace*{-0.75cm}

Consequently, both experts were asked to discriminate between severe and non-severe cases, for which results are shown in Table~\ref{tab:expert_severity_confusion}, with purities of $0.73$ and $0.87$ respectively.

\begin{table}[!h]
	\begin{center}
			\begin{tabular}{*{4}{|c}|}
				\hline
				\multicolumn{2}{|c|}{\multirow{2}{*}{Severe case}} & \multicolumn{2}{c|}{Expert 1 \cellcolor[gray]{0.6}} \\ \cline{3-4}
				\multicolumn{1}{|c}{} & \multicolumn{1}{c|}{} & \cellcolor[gray]{0.8} Yes & \cellcolor[gray]{0.8} No \\ \hline
				\cellcolor[gray]{0.6} & \cellcolor[gray]{0.8} Yes & 8 & 2  \\ \cline{1-4}
				\cellcolor[gray]{0.6} & \cellcolor[gray]{0.8} No & 6 & 14 \\ \cline{1-4}
				\multicolumn{1}{c}{\multirow{-4}{*}{Model}} & \multicolumn{3}{c}{}
			\end{tabular}
			\begin{tabular}{*{4}{|c}|}
				\hline
				\multicolumn{2}{|c|}{\multirow{2}{*}{Severe case}} & \multicolumn{2}{c|}{Expert 1 \cellcolor[gray]{0.6}} \\ \cline{3-4}
				\multicolumn{1}{|c}{} & \multicolumn{1}{c|}{} & \cellcolor[gray]{0.8} Yes & \cellcolor[gray]{0.8} No \\ \hline
				\cellcolor[gray]{0.6} & \cellcolor[gray]{0.8} Yes & 7 & 3  \\ \cline{1-4}
				\cellcolor[gray]{0.6} & \cellcolor[gray]{0.8} No & 1 & 19 \\ \cline{1-4}
				\multicolumn{1}{c}{\multirow{-4}{*}{Model}} & \multicolumn{3}{c}{}
			\end{tabular}
		\vspace*{-0.5cm}
		\caption{Severe case confusion matrices.}
		\label{tab:expert_severity_confusion}
	\end{center}
\end{table}

\paragraph*{Association with gold-standard questionnaire.}
We now summarize our results on how the learned phenotypes associate with
responses to the WERF EPHect questionnaire. In general, severity and quality of
life indicators of endometriosis as specified by WERF standards align well with
how our model discriminates patients. 

Overall, phenotype 0 (severe phenotype) is associated with a heavier burden of
disease. It shows correlation with a number of comorbidities such as anxiety
(odds ratio, OR=1.62), depression or mood disorders (OR=1.57), migraine
(OR=2.27), high blood pressure (OR=1.69), and polycystic ovary syndrome 
PCOS (OR=1.81). It is also associated with painful bladder problems like
interstitial cystitis (OR=2.28), which did not occur in patients in other
phenotypes. Symptoms of these comorbidities are salient in the obtained
posteriors (see for example, how the severe phenotype is characterized by painful urination).
Furthermore, women with endometriosis are known to exhibit these comorbidities:
anxiety, depression, and other mood disorders \citep{j-Pope2015}, migraines
\citep{j-Yang2012}, high blood pressure \citep{j-Mu2017}, PCOS
\citep{j-Holoch2014}, and painful urination/interstitial cystitis
\citep{j-Chung2005}.  

Participants assigned to phenotype 0 also show an important reduction in
quality of life, as compared to those in other phenotypes. Patients within this subtype were 3.25 more likely to rate their health as poor.
Patients within the severe phenotype are more likely (OR=1.70) to
report problems regarding going to work or carrying out activities of daily
living as well. These participants had also higher odds of experiencing limitations
with vigorous activities such as running (OR=2.06), moderate activities
like housework (OR=3.45), lifting (OR=3.86), climbing even one flight
of stairs (OR=4.42), bending (OR=3.52), walking just one block
(OR=5.47), dressing, as well as bathing (OR=10.67). Problems with
productivity and reduced quality of life among endometriosis patients has been previously reported \citep{j-Shabanov2017}. 

The surgical burden of the disease is most evident for those subjects
assigned to phenotype 0. Participants in this group had significantly higher number of
laparoscopies (mean 1.67) as compared to those in phenotypes 1 and 2 (mean
1.36).  Consistent with the existing literature \citep{j-Ballard2006},
patients also reported seeing several doctors before diagnosis, with phenotype 0 being
associated with the highest number of doctors seen (mean=6.16), versus those in other phenotypes (mean=4.59).


Several other associations were consistent with current disease knowledge. The
severe phenotype had 10.86 times higher odds of having mild endometriosis
(Stage 2) as compared to phenotypes 1 and 2, and the mild phenotype had 0.48
times lower odds of having a surgery where no endometriosis was found as
compared to phenotypes 0 (severe) and 1 (moderate). While these results might
seem counterintuitive, they confirm the lack of correlation between patient
experience and existing staging of disease as discussed in the literature
 \citep{j-Vercellini2007}. In addition, the odds of having an
extremely regular period were 0.47 times lower in phenotype 0 (severe
phenotype) compared to others. Such menstrual irregularity has been
shown to be associated with endometriosis before \citep{j-Signorello1997}. 

Finally, associations were also found that may indicate differences in the
etiology of the disease among the different phenotypes. Those in phenotype 2 had
2.50 times higher odds of having a sister with endometriosis than others. Many
different causes of endometriosis have been proposed including heritable
tendencies, but the exact etiology remains unclear
\citep{ic-Cramer2004}.
The set of proposed underlying causes of the diseases highlights the need for
reducing the heterogeneity of clinical presentation of symptoms, and our
approach seems promising for reducing this heterogeneity.   
Finally, we found no significant correlations between phenotypes and age, race, or time-to-diagnosis.

\section{Conclusion} 
This paper contributes to research in digital phenotyping from self-tracking
data. Our joint modeling of multiple types of self-tracking variables through
mixed-membership models show that we can produce robust, clinically meaningful
groupings of self-tracked variables. These phenotypes, along with participant
clustering, suggest novel findings about disease.

In the case of endometriosis, a particularly enigmatic condition with a dire need for phenotyping and subtyping, our methods identified three clusters of patients roughly grouped by
the severity of their condition. Further, clinically meaningful novel
associations beyond what is currently known about the disease were identified.
Endometriosis phenotypes are necessary as a first step towards a better understanding of the pathophysiologic mechanisms of the disease, which will lead towards better treatment and management of patients based on learned phenotypic characteristics.

Future work should include modeling the temporality of signs and symptoms of
endometriosis, particularly since it is estrogen dependent and thus linked to
the menstrual cycle. Nevertheless, the analysis in this study already shed
novel insight and demonstrates the value of patient-generated data in medical
research.

\acks{The authors thank the study participants and acknowledge the many endometriosis advocacy groups that helped in recruiting them. This work was supported by the Endometriosis Foundation of America, the National Science Foundation award SCH 1344668, and the National Library of Medicine award T15 LM007079. We also thank Dr.\ Shadi Safar Gholi and Dr.\ Arnold Advincula for their time and expertise.}

\bibliography{mlhc}

\clearpage
\appendix
\section*{Appendix A.}
\label{sec:appendix_A}
Vocabulary for each tracked question:

\begin{itemize}
\item \textbf{Where is the pain}: bones\_pain, cervix\_pain, deep\_vagina\_pain, diaphragm\_pain, head\_pain, inner\_thighs\_pain, intestines\_pain, joints\_pain, left\_arm\_pain, left\_breast\_pain, left\_leg\_pain, left\_lower\_back\_pain, left\_outer\_hip\_pain, left\_ovary\_pain, left\_pelvis\_pain, left\_ribs\_pain, left\_shoulder\_pain, left\_side\_abdomen\_pain, legs\_pain, lower\_back\_pain, lower\_chest\_pain, neck\_pain, pelvis\_pain, rectum\_pain, right\_arm\_pain, right\_breast\_pain, right\_leg\_pain, right\_lower\_back\_pain, right\_outer\_hip\_pain, right\_ovary\_pain, right\_pelvis\_pain, right\_ribs\_pain, right\_shoulder\_pain, right\_side\_abdomen\_pain, upper\_abdomen\_pain, upper\_chest\_pain, uterus\_pain, vagina\_entrance\_pain, whole\_abdomen\_pain
\item \textbf{Describe the pain}: aching\_pain, burning\_pain, cramping\_pain, deep\_pain, dull\_pain, nauseating\_pain, pressure\_pain, pulling\_pain, pulsating\_pain, radiating\_pain, sharp\_pain, shooting\_pain, stabbing\_pain, throbbing\_pain, twisting\_pain
\item \textbf{How severe is the pain?}: mild\_pain, moderate\_pain, severe\_pain
\item \textbf{What are you experiencing}: allergies, asthma, blurry\_vision, chest\_pressure, dizziness, eczema, fatigue, fever, headache, hives, hot\_flash, itchy, mentally\_foggy, noise\_sensitivity, numbness, rash, ringing\_in\_ears, sinus\_congestion, sweaty, swelling, touch\_sensitivity
\item \textbf{How severe is the symptom}: mild\_symptoms, moderate\_symptoms, severe\_symptoms
\item \textbf{Describe the flow}: heavy\_flow, light\_flow, medium\_flow
\item \textbf{What kind of bleeding}: breakthrough\_bleeding, clots, no\_bleeding, spotting
\item \textbf{Describe GI/GU system}: blood\_in\_stool, cant\_urinate, constipation, diarrhea, endo\_belly, frequent\_urination, gas, heartburn, mouth\_sores, nausea, painful\_bowel\_movement, painful\_urination, stomach\_upset, uncomfortably\_full, vomiting
\item \textbf{How severe is it}: mild\_GI, moderate\_GI, severe\_GI
\item \textbf{Describe sex}: avoided\_sex, bleeding\_from\_sex, no\_sex, painful\_after\_sex, painful\_during\_sex, sex\_felt\_good
\item \textbf{Activities}: climb\_stairs, eat, get\_dressed, get\_out\_of\_bed, have\_sex, housework, jump, kneel, lie\_down, lift, no\_trouble, prepare\_food, run, shop, shower, sit\_down, sleep, socialize, stand, stretch, use\_toilet, walk, work
\item \textbf{How was your day?}: bad\_day, good\_day, great\_day, manageable\_day, unbearable\_day
\item \textbf{Medications/hormones taken}: adrenergic\_agonists, amphetamine, analgesic, analgesic/narcotic, analgesic/nsaids, analgesic/opioids, anesthetic, anorectic, anti-inflammatory, antiacid, antiacid/nsaids, antibiotics, anticholinergic, anticoagulant, anticonvulsant, antidepressant, antidiabetic\_medication, antidiarrheal, antiemetic, antihistamine, antihypertensive, antipsychotic, antispasmodic, antispasmodic/sedative, anxiolytic, anxiolytic/anesthetic/muscle\_relaxant, barbituate, barbituate/analgesic, beta\_blocker, bronchodilator, calcium\_channel\_blocker, cough\_medicine, decongestant, diuretic, dopamine\_agonist, estrogen, estrogen/progestin, gonadotropin-releasing\_hormone\_agonist, gonadotropin-releasing\_hormone\_antagonist, hormone\_based\_chemotherapy, hormone\_replacement\_therapy, human\_chorionic\_gonadotropin, human\_follicle\_stimulating\_hormone, laxative, muscle\_relaxant, narcotic, narcotic/nsaids, neuropathic\_pain\_medication, no\_med\_hormones, noclass, nonbenzodiazepine\_hypnotic, nsaids, opioids, progestin, sedative, statin, steroid, stimulant, thyroid\_hormones, topical\_anti-tumor\_medication, triptan, vasoconstrictor, vitamin\_a\_derivative
\end{itemize}

\end{document}